# HELENE: An Open-Source High-Security Privacy-Preserving Blockchain Based System for Automating and Managing Laboratory Health Tests

Gabriel Fernández-Blanco[a], Pedro García-Cereijo[a], David Lema-Núñez[a], Diego Ramil-López[a], Paula Fraga-Lamas[a,b,e], Leire Egia-Mendikute[c], Asís Palazón[c,d], Tiago M. Fernández-Caramés[a,b]

[a]*Department of Computer Engineering, Faculty of Computer Science, Universidade da Coruña, A Coruña, 15071, Spain*
[b]*Centro de Investigación CITIC, Universidade da Coruña, A Coruña, 15071, Spain*
[c]*Cancer Immunology and Immunotherapy Laboratory, Center for Cooperative Research in Biosciences (CIC bioGUNE), Basque Research and Technology Alliance (BRTA), Bizkaia Technology Park, Building 801A, Derio, Spain*
[d]*Ikerbasque, Basque Foundation for Science, Bilbao, Spain*
[e]*Corresponding author: Paula Fraga-Lamas, paula.fraga@udc.es*

**Abstract**

In the last years, especially since the COVID-19 pandemic, precision medicine platforms emerged as useful tools for supporting new tests like the ones that detect the presence of antibodies and antigens with better sensitivity and specificity than traditional methods. In addition, the pandemic has also influenced the way people interact (decentralization), behave (digital world) and purchase health services (online). Moreover, there is a growing concern in the way health data are managed, especially in terms of privacy. To tackle such issues, this article presents a sustainable direct-to-consumer health-service open-source platform called HELENE that is supported by blockchain and by a novel decentralized oracle that protects patient data privacy. Specifically, HELENE enables health test providers to compete through auctions, allowing patients to bid for their services and to keep the control over their health test results. Moreover, data exchanges among the involved stakeholders can be performed in a trustworthy, transparent and standardized way to ease software integration and to avoid incompatibilities. After providing a thorough description of the platform, the proposed health platform is assessed in terms of smart contract performance. In addition, the response time of the developed oracle is evaluated and NIST SP 800-22 tests are exe-cuted to demonstrate the adequacy of the devised random number generator. Thus, this article shows the capabilities and novel propositions of HELENE for delivering health services providing an open-source platform for future researchers, who can enhance it and adapt it to their needs.

## 1. Introduction

Although the world has entered into a post-pandemic era after the global COVID-19 outbreak, important healthcare bottlenecks remain, including challenges in primary care, medical data privacy and sustainability of the different National Health Services [1]. This situation was aggravated and fueled by the emergence of SARS-CoV-2 variants, vaccination campaigns (e.g., booster timing, protection against variants, immunity passport) and drug discovery efforts (e.g., therapeutic neutralization against emerging variants, vaccine and antibody development) [2, 3]. In addition, the saturation of fully centralized health services provided by National Health Services was an important problem, in part as a result of sequential infection waves and intense testing and vaccination campaigns [4]. Such a situation resulted in non-optimal primary care for COVID and non-COVID patients and on urgent medical necessities that still need to be addressed to prevent future problems [5].

Moreover, healthcare users (patients) are increasingly demanding better quality and more speed in the provision of health services. In such a context, data privacy is an emerging issue that has been traditionally handled by certain entities (e.g., healthcare services, pharmaceutical companies, governmental agencies, insurance companies) in a way that is perceived by many users as non-transparent. This is particularly relevant in the context of genomic data, but the threat is extending to other areas of health in the post-pandemic era (e.g., vaccination and infective status, level of antibodies, emerging variants) [6]. A clear example is the requirement of immunity passports to demonstrate previous vaccination in order to carry out basic activities such as traveling and networking, even when the level of neutralizing antibodies for each individual at a given time point is unknown and, as of writing, current COVID vaccines still do not prevent community transmission [7].

Regarding data protection needs, blockchain has arisen in the last years as a disruptive technology that allows users to keep track of who uses their data [8, 9], to decide how such data are managed [10], to contribute easily with their data to research [11] or to pharmaceutical development projects (i.e., vaccines against emerging variants and epidemiology studies [12], vaccine passports [13]), and even to receive a reward for transferring health data to companies (i.e., vaccine developers) [14].

To tackle the previously mentioned issues, this article presents HELENE (**H**igh-s**E**curity privacy-preserving digital hea**L**th s**E**rvice platform for the post-pa**N**demic **E**ra), which is an open-source blockchain-based smart digital health platform that is aligned with the digitalization needs of the post-pandemic era, considering aspects related to health data privacy and traceability. Specifically, HELENE provides a DApp (Decentralized Application) that is aimed at satisfying two main requirements:

1. **Competitive patient request provision**: HELENE provides a bid system to guarantee the best service to the patient. Thus, it creates an ecosystem where laboratories compete with each other to get their offers accepted by patients.
2. **High-security health test data management**: the Test Results (TRs) related to a patient from test samples are securely managed, providing access only to authorized parties.

The following are the main contributions of this article:

- A thorough description is provided on the design and implementation of what to the knowledge of the authors is the first blockchain-based health test platform aimed at performing direct-to-consumer immunoassays, which is available as open-source software [15].

- Novel strategies are proposed for delivering decentralized health test services by independent laboratories, which compete in a transparent way to provide the best product/service in terms of price, quality or delivery time. In addition, since the platform allows for controlling patient private data, it enables implementing new strategies for incentivizing data sharing among the participating entities.

- The development and evaluation of a decentralized external oracle is detailed. Such an oracle relies on a explicitly-devised pseudo-random number generator, whose security is evaluated through the set of statistical tests defined by NIST SP 800-22.

- In order to carry out realistic tests, the output of a real test machine is modeled, implemented and integrated with the health platform. The model follows a real data distribution from COVID-19 test samples that the authors of the article collected in 2020 and 2021.

The rest of this article is structured as follows. Section 2 reviews the background and previous work, covering the evolution from traditional laboratory testing procedures to the current challenges in health data, as well as concepts related to Distributed Ledger Technologies, and blockchain-based systems for lab tests and pharmaceuticals. Section 3 describes the design of the proposed system, including the communications architecture, the blockchain-managed lab test procedure and the main interactions. Section 4 describes the lab test procedure and the performed test machine modeling. Section 5 describes the implementation of the proposed architecture, detailing the bidding system smart contracts and their standardization, as well as the used decentralized storage, Oracle Nodes, secure pseudo-random number generation API, the lab test machine simulator, and the solution front-end. Finally, Section 6 presents the conducted experiments and Section 7 is devoted to conclusions.

## 2. Background and previous work

*2.1. Traditional laboratory test procedure*

*2.1.1. Overall procedure*

The procedure for conducting the laboratory tests begins with patients submitting a request. Upon approval, patients are instructed to visit the facility to provide the required biological samples, such as plasma or saliva. Alternatively, samples may be collected by healthcare professionals at clinical sites and transported to the laboratory under appropriate conditions. Once the samples arrive, they are processed according to standardized protocols, analyzed using specialized equipment, and the results are reviewed by qualified personnel. Test outcomes are reflected in reports that are delivered to patients electronically, via email or through a secure online system.

*2.1.2. Immunoassays*

To date, most immunoassays that determine seroconversion and measure antibody responses are based on enzyme-linked immunosorbent assay (ELISA), including automated chemiluminescent variants used in the clinic (i.e., CLIA). ELISAs have an acceptable specificity and sensitivity profile for performing large epidemiological studies, but their sensitivity in the context of, for instance, SARS-CoV-2 serology could be improved. A key limitation of ELISA is the need for individual plates/wells for each antigen or antibody class to be tested. The lack of multiplexing capacity also results in the need for more material, which is usually not sustainable and damages the environment (i.e., tips, plastic plates, packaging, gloves). Moreover, in ELISAs, the antigen is immobilized to the plate, which can hide epitopes or increase the background noise. For these reasons, ELISAs are not well suited to detect low antibody titers and often give undetermined values that are close to the cut-off value, leading to difficult interpretation of results.

Another established approach to measure seroconversion (i.e., to carry out antibody tests) or the presence of antigen (i.e., through antigen tests) is based on lateral flow assay technology. This approach has the main advantage of genuine point of care capability and timely results in a high-incidence scenario. The main limitation of this approach is that the detection threshold is not well suited to detect small amounts of antibodies or viral antigens. In addition, it has limitations of self-testing associated with incorrect sample collection techniques by the end-user.

For these reasons, lateral flow assays are not suitable for measuring antibody titers in a quantitative manner, or detecting infected individuals with low viral load or emerging variants with mutations that escape the limit of detection of tests designed to identify, in the case of SARS-CoV-2, the original Wuhan variant. The supply and quality of lateral flow assays is not always guaranteed, and their deployment at National Health Centers at the peak of infection waves often results in a collapse of the service, delays, or limited impact on containing infection. As a consequence, it is clear that the digitalization of these services can alleviate their saturation.

Besides the previously mentioned issues, as we enter into the post-pandemic era, important bottlenecks remain. These include the need for technologies that support the identification of infected individuals (including asymptomatic and emerging variants), vaccination follow-up (e.g., booster timing, protection against variants, immunity passports) and drug discovery (e.g., therapeutic neutralization against emerging variants, antibody development). In this context, HELENE harnesses a recently developed precision serology platform previously developed by the authors that detects the presence of antibodies and antigens with superior sensitivity and specificity at a fraction of the time and cost compared to traditional methods (i.e., ELISA, lateral flow) [16]. Such a solution is based on an immunological technique called cytometric bead array, which leverages multiplexing to also achieve an important reduction of the required amount of plastic (tips, plates, gloves) [17].

*2.2. Main challenges related to health data privacy, anonymity and security*

One of the challenges currently faced by health providers, practitioners and users is how to collect, validate and store health information in a secure, flexible, scalable and low-cost way, while respecting data privacy and anonymity [18]. In addition, it is necessary to avoid untrusted parties while providing transparent and trustworthy data sources, whose information comes from people that can differ in aspects like age, ethnicity, education or geographic location [19].

Although technologies like data crowdsourcing can enable providing novel smart health services (e.g., diagnosis services, patient monitoring, disease control solutions) [20], many patients feel that they lose control over their health data when they provide them to third parties, which may speculate with them to get an economic profit. In addition, traditional database-based approaches, where health information is stored in a centralized way, are prone to cyberattacks, data leaks, data losses and usually suppose a Single Point of Failure (SPoF) [21]: if the access to the database fails (e.g., due to a cyberattack, a malfunction, maintenance), the whole system usually stops working.

There are also interoperability problems among the different available technological health platforms, which often operate as independent proprietary software and hardware solutions that are incompatible among them and that are difficult to scale [22]. Such solutions impede the creation of novel end-to-end patient-centered systems for research and healthcare. This is related to the fact that the involved stakeholders (e.g., National Health services, insurance companies, software developers) have to share information that can be incomplete, fragmented, not available at the point of care, or that it is difficult to access [23]. There are specific standards for exchanging health data (e.g., Health Level Seven (HL7) Fast Healthcare Interoperability Resources (FHIR), HL7 Clinical Document Architecture (CDA), ISO13606, openEHR, CDISC Operational Data Model (ODM)), but, unfortunately, their practical implementation is not straightforward, requiring the dedication of a significant amount of time to data mapping and additional interface adaptations [24].

*2.3. On the Use of Blockchain and Distributed Ledger Technologies*

Currently, there are many domains where traditional centralized systems have not been able to provide the required data security, data integrity and data privacy [25]. As it was mentioned in the previous Section, the healthcare domain still struggles with fragmented medical records, unauthorized data accesses and inefficiencies in sharing information across multiple providers. These issues are inherent to traditional systems, where data are stored on centralized servers. This introduces vulnerabilities such as SPoFs or Denial of Service (DoS), leading to potential unavailability and recurrent cyber-attacks that expose patient data [26].

To address these challenges, disruptive technologies like blockchain have emerged. A blockchain is a type of Distributed Ledger Technology (DLT) that manages transactions by linking them in the form of a chain through cryptographic hashes [27]. Some blockchains allow for automating certain actions through smart contracts, which are actually programs that are stored on a blockchain and that are executed by the blockchain participants or self-executed when certain conditions are met.

A blockchain is deployed on decentralized communications architectures that provide secure data exchanges, while guaranteeing data privacy, integrity, accountability, transparency and resiliency against cyberattacks [28]. Such a decentralized nature enables increasing operational efficiency and removes the need for unnecessary intermediaries that often incur in delays and additional costs when providing healthcare services. The key advantage of a blockchain is that it provides the previously mentioned benefits between parties that do not necessarily trust each other. This is the case of patients that do not trust certain entities that collect their health information and that may end up selling it to third parties.

As the post-pandemic era has arisen unprecedented questions about how patients can effectively share their data, about how to prevent document forgery [29] or on the sustainability of current procedures or the management of tests results, a blockchain and other decentralized technologies can be leveraged to effectively address these new challenges through the development of DApps.

*2.4. Blockchain-based systems for lab tests and pharma*

Direct-to-consumer health testing provides people direct access to their health information without necessarily involving a healthcare provider or health insurance company in the process. Many companies currently offer direct-to-consumer genetic tests (e.g., 23andme, MyHeritage) and this trend is expanding into, for instance, COVID-19 testing (e.g., Everlywell). There is a privacy problem associated with health data collected by these companies, which could be sold to pharmaceutical or insurance companies without the control of the end-user [30, 31].

DLTs provide a good solution for managing health data ownership and health data transactions in a private, transparent and decentralized manner [32, 33, 34]. Specifically, blockchain allows for evolving towards a collaborative economy where peers can exchange money and services through crowdsourcing data sharing systems [35, 36]. In fact, the World Economic Forum (WEF) forecasts that, by 2027, 10% of the global gross domestic product will be stored on blockchains [37].

Due to the previously mentioned benefits, blockchain has been used by many industries, like in the automotive industry [38], in logistics [39], in the agri-food industry [40], in transportation [41], by energy companies [42] or in Industry 4.0/5.0 smart factories [43, 44]. The healthcare industry has also made use of blockchain. For example, in [45] the authors describe an application for the pharma supply chain that makes use of sensors and a blockchain to preserve data integrity and provide public accessibility to temperature records, which is essential to guarantee the quality of the transported medical products. Thus, in the devised system, every shipped parcel embeds a sensor that collects environmental information and sends it to a blockchain where a smart contract is executed to determine whether the sensor values are within the allowed storage range. Another interesting blockchain-based smart health system is presented in [46], where the authors detail a platform for clinical trials and precision medicine. It is also worth mentioning the work presented in [47], which describes the design of a general architecture based on blockchain for providing smart healthcare services.

In contrast to the previously mentioned state of the art, this article presents a sustainable direct-to-consumer platform that allows patients to access health data in a private and secure manner supported by blockchain technology and that provides a transparent auction-based mechanism used by the health providers to compete in an incentivized way. Moreover, the proposed platform has been designed to support the use of specific lab test machines able to determine the level of antibodies against SARS-CoV-2 present in dry blood spot samples and can establish their neutralization capacity against emerging variants. The employed high-sensitivity immunoassays address the previously described challenges and serve as a demonstrator of decentralized digital health in the post-pandemic era, with direct applications in COVID-19 and beyond.

Thus, direct applications of the developed system include the secure and sensitive determination of the level of antibodies, which has relevant implications in the rationalization of vaccine boosters, immunity passports, national epidemiology studies and pharmaceutical development. Moreover, the provided blockchain-enabled digital solution follows the Industry 5.0 principles [48], so it is resilient and human-centered: it protects patient privacy (e.g., during the purchase of sample collection kits or when consulting TRs) by using high-security mechanisms, while enabling patients to determine how health data should be managed. Thus, the developed system guarantees secure data exchanges between patients and healthcare service providers, providing trustworthiness and transparency for patient data sharing. Furthermore, data exchanges among the involved stakeholders are performed in a standard way to ease software integration and to avoid incompatibilities.

Finally, it is worth noting that the proposed blockchain-based platform implements a novel mechanism to distribute service requests among potential analytical laboratory providers, which compete in a transparent way to provide the best product/service in terms of price, quality or delivery time. Thus, the platform is able to make use of mechanisms to incentivize the use of the system and thus reward users depending on how they decide to manage their health data.

## 3. Design of the system

### 3.1. Blockchain technology

Among the multiple available blockchain technologies, it is necessary to distinguish public, private and consortium/federated blockchains [43]:

- Public blockchains. This type of networks allow unrestricted participation, so no approval is needed to join to them. Users can both

publish and validate transactions, and miners (i.e., those responsible for validating blockchain transactions) are typically rewarded for their validation efforts. Moreover, public blockchains are particularly suitable for industries requiring high transparency or scenarios involving large-scale consumer-device interactions.

- Private blockchains. In these networks, access is controlled by a designated entity, which dictates aspects such as mining incentives, network participation and validator selection. Since a single entity governs the blockchain, the network cannot be considered as truly decentralized. Instead, it functions more like a secure distributed database. This structure can be advantageous in scenarios where participants are well-known and auditing processes are necessary.

- Consortium or federated blockchains. These blockchains are managed by multiple entities, which collectively regulate network access and participant permissions. Typically, a predefined group of nodes executes the consensus mechanism, enhancing privacy and expediting transaction validation. Such a model is beneficial for collaborations that require transaction validation and data exchange. For instance, healthcare entities may each maintain a validation node and a transaction is recorded only when a quorum of nodes approves it.

In addition, blockchains can be classified based on other relevant characteristics:

- Based on permissions:
  - Permissionless blockchains: all users have equal access to perform actions on the blockchain, eliminating the need for permission management.
  - Permissioned blockchains: these blockchains impose restrictions on which users can carry out transactions.

- Based on incentives:
  - Tokenized blockchains: their transactions and incentives are linked to tokens that participants exchange or earn. For instance, Bitcoin operates using bitcoins, while Ethereum relies on Ether.

- Non-tokenized blockchains: these blockchains do not require a specific virtual currency.

- Based on operation mode:

  - Logic-oriented blockchains: this type of blockchains support the execution of complex logic, with smart contracts being their most well-known application, although they also enable using other forms of automation.
  - Transaction-oriented blockchains: these blockchains are specifically designed to track digital assets.

Considering the previous alternatives, for the field of healthcare service delivery, private or consortium permissioned tokenized logic-oriented blockchains are more appropriate because they ensure data privacy, regulatory compliance, controlled access, rewarding mechanisms and automation while still providing blockchain benefits like security and transparency. Table 1 compares the main features of the most relevant private and consortium logic-oriented blockchains.

Table 1: Main features of the most relevant private and consortium logic-oriented blockchains.

| Feature | Hyperledger Fabric | R3 Corda | Quorum | Ethereum |
|---|---|---|---|---|
| Type | Consortium | Consortium | Private | Public/Private |
| Consensus Protocol | Modular (Kafka/Raft) | Notary Service | RAFT/IBFT | Proof-of-Work /Proof-of-Authority (Clique) |
| Transaction Speed | 1-5 s | Instant | Instant | 15 s-5 min |
| Smart Contracts | Chaincode (Go/Java) | Kotlin/Java | Solidity | Solidity/Vyper |
| Node Permissioning | Certificate-based | Legal Identity | Smart Contracts | Permission Flags |

Among the blockchains compared in Table 1, Ethereum [49] was selected for the solution presented in this article due to its smart contract functionality, security, interoperability and decentralization. Specifically:

- Its smart contracts allow for creating a logic-oriented blockchain able to automate healthcare service delivery workflows.

- It is a tokenized blockchain that allows for implementing incentivization mechanisms to foster the participation of users and entities.

- It provides high-enough transaction speed for the proposed application (i.e., for healthcare service delivery is not necessary to be able to perform thousands of transactions per second).

- In contrast to the other blockchains compared in Table 1, it is more flexible, since it can be easily deployed in both private and public networks, which can be useful for certain healthcare applications. For instances, for an application aimed at providing laboratory health tests where private information is exchanged, a private blockchain is more appropriate. However, for a vaccination passport solution, whose information may be publicly shared with many potential entities, a public solution is a better fit.

- Ethereum provides different standards (analyzed in the next subsection) that allow for the secure exchange of information among healthcare providers and entities and for creating incentives to reward the participants.

*3.2. Ethereum Standardization*

One of the problems related to health is data exchange. This is also a problem for DLTs like blockchain, so the different networks developed initiatives to create standards. In the case of Ethereum, its standards are known as ERCs (Ethereum Request for Comments) [50]. An ERC provides a description of general functionalities and use cases, along with the required methods and parameters needed to comply with specified conventions.

The blockchain-based solution described in this article, is a decentralized Ethereum-based system that involves payments between patients and laboratories, thus requiring a specification capable of granting secure, trustworthy and efficient payment exchanges between parties. Moreover, it must provide regulatory authorities with the ability to override transactions in cases where the system is misused or abused, and support an in-chain documentation registry.

As of writing, three ERCs are appropriate for providing the previously described features (i.e., security, regulatory oversight and efficient transaction capabilities): ERC-20, ERC-1400 and ERC-3643. Each of these standards provide a different set of functionalities and trade-offs:

- ERC-20 is one of the most widely adopted Ethereum standards, being used for a large number of tokens [51]. It is known for its simplicity and ease of implementation [52], allowing for basic token transfer operations or the delegation of such a capability to a trusted party. However, these characteristics result in a limited set of functionalities:

ERC-20 primarily supports basic transfers and lacks a straightforward mechanism for regulatory authorities to reverse unlawful transactions. Moreover, it does not support in-chain document registry.

- ERC-1400 was developed to replicate the stock exchange market on the Ethereum blockchain [53], where tokens compliant with this standard grant specific privileges to certain token holders within on-chain corporations, referred to as partitions. Similarly to ERC-20, there is a role (operator) for performing transfers with the approval of the token holder. An operator can be defined per partition or for the entire contract. However, the operator role is insufficient to simulate regulatory authorities, as it has limitations on the number of tokens it can transfer and requires prior approval from the token holder. The controller role, in contrast, is better suited for such a purpose, acting as an imposed authority with the ability to force transfers of any amount without requiring consent. Finally, this standard also supports in-chain document metadata storage, a feature of future interest for HELENE as it enables the creation of on-chain records linked to off-chain documents like health reports or TRs.

  In Figure 1, a diagram is provided with the different actors involved in the ERC-1400 functioning. Two controllers are on top as overseers of the whole system. Right below there is an ERC-1400 contract with two operators assigned by the instance owner. Finally, at the bottom, there is a set of partitions of a previous ERC-1400 contract, where each partition has one operator except the one on the right. In a certain way, a partition acts as an independent contract within the general ERC-1400 contract.

- ERC-3643 is designed to replace ERC-1400, addressing its lack of key features necessary for compliance with various legal systems, such as adaptability to different jurisdictions [54]. ERC-3643 introduces KYC (Know Your Customer) interfaces [55], enabling a trusted third party to verify participants and enforce legal compliance, including restrictions on transactions between different countries [54, 55]. In addition, ERC-3643 implements batch operations to optimize resource usage. The documentation registry is replaced with an on-chain automated system, eliminating the reliance on external documents. Furthermore, compliance mechanisms in ERC-3643 allow the definition of regulatory authority addresses. Figure 2 illustrates the different sections of the standard. At the top, each actor's key is verified by a trusted party, which issues an identity claim stored on-chain. These identity claims, along with compliance claims specifying enforceable restrictions, are used to verify transactions between parties, as shown at the bottom right. This verification applies to both regular and batch operations.

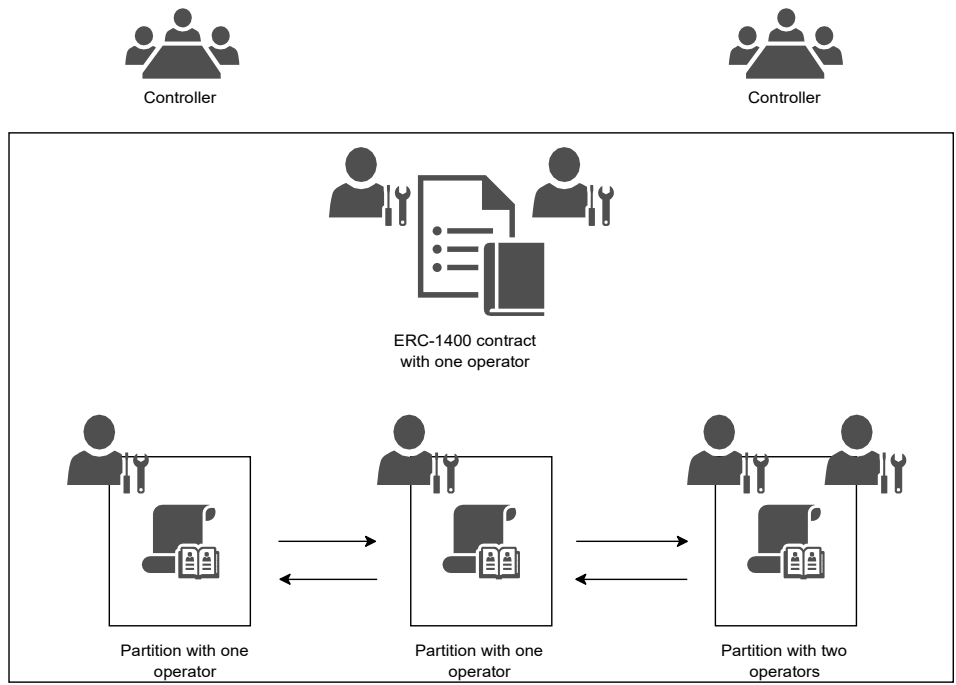

Figure 1: ERC-1400 entity interaction.

*3.3. Communications architecture*

To improve the data privacy, security and resilience of traditional centralized cloud-based solutions, HELENE makes use of the blockchain-based architecture depicted in Figure 3. Such an architecture stores the transactions on a blockchain, which also runs smart contracts that automate multiple involved processes. Information is stored in a decentralized way: each smart health service provider (e.g., an analytical laboratory) runs a local storage node (e.g., based in OrbitDB/IPFS) [56, 57], which allows for synchronizing the stored TRs in a secure and distributed way. Thus, every TR stored by a health provider will be automatically replicated in all the providers that participate in the healthcare solution, so all of them will have almost instant access to the data stored by any provider (if they are provided with the neces-sary permissions to access them). Similarly, patients can access their health data securely either directly through the blockchain or by using a web-based front-end.

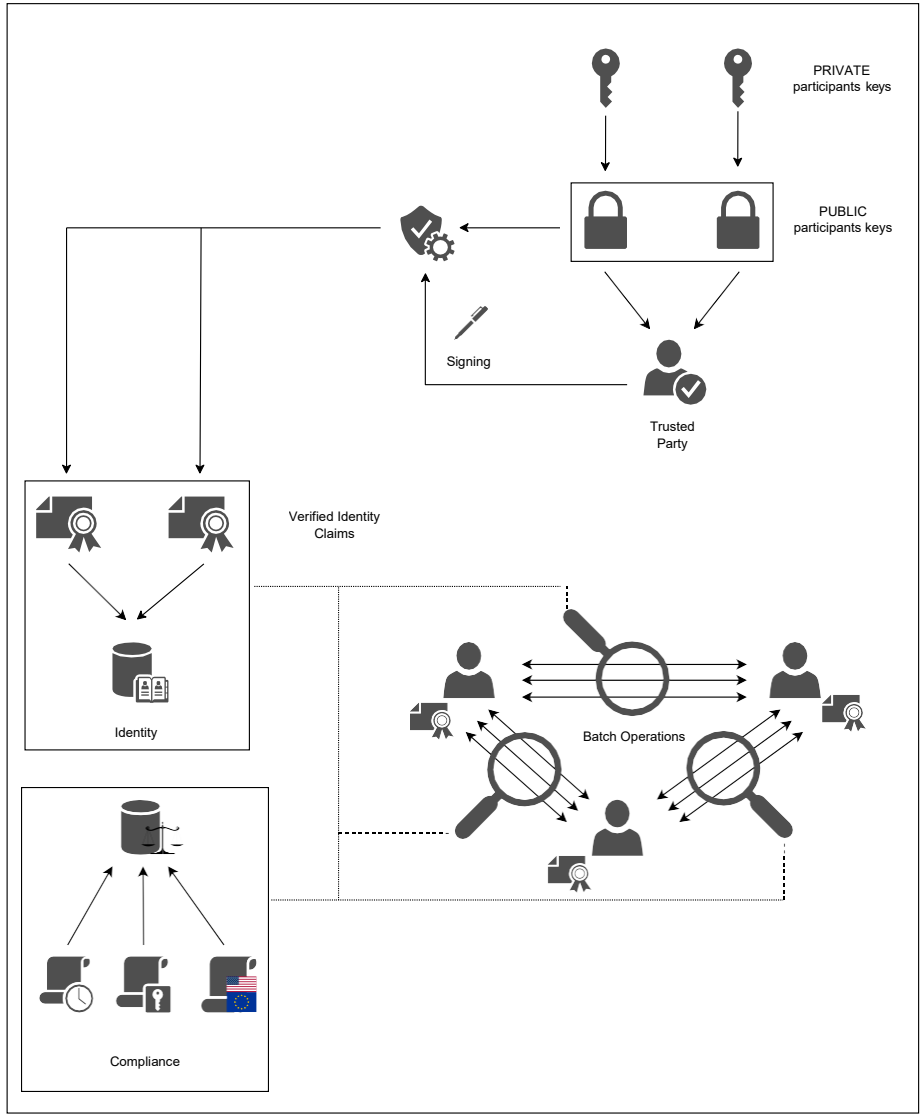

Figure 2: ERC-3643 entity interaction.

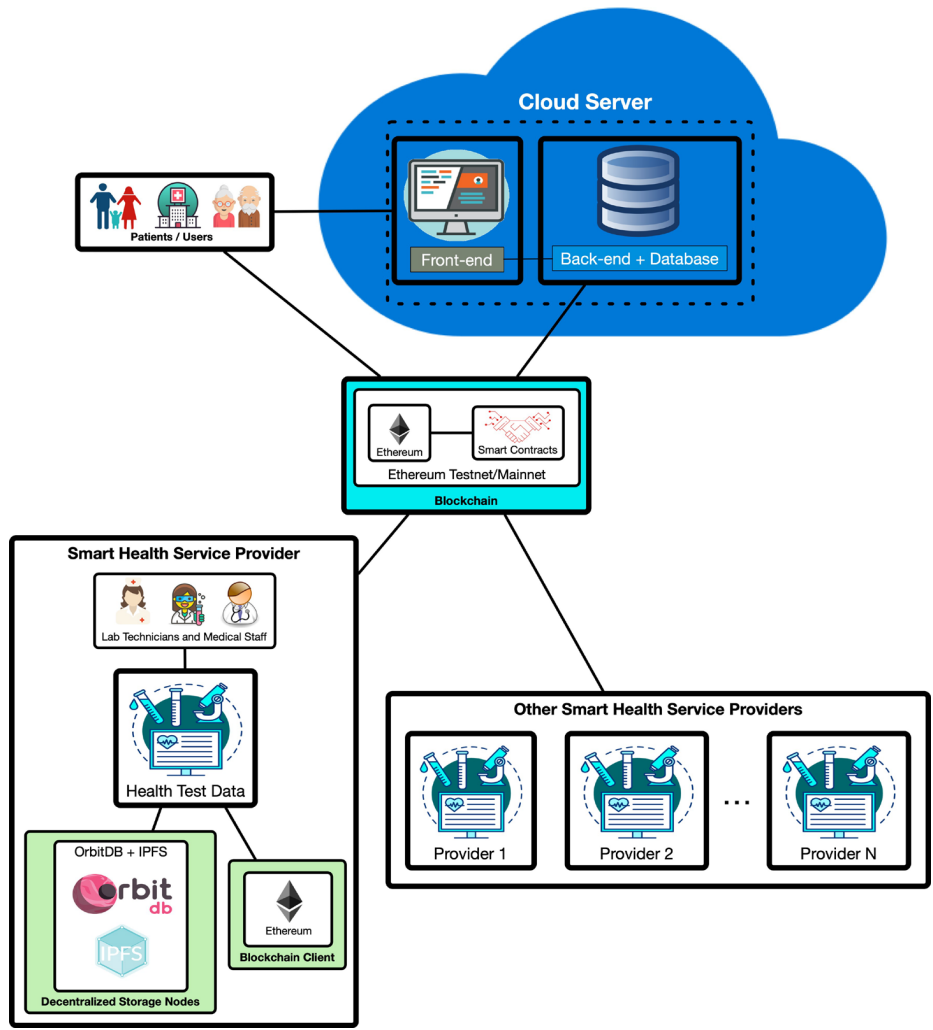

Figure 3: HELENE's blockchain-based communications architecture.

## 3.4. Overview of the proposed blockchain-managed lab test procedure

The interactions among the main entities involved in the inner workings of the proposed solution are illustrated in Figure 4. Such a figure shows a UML sequence diagram where first, a patient (or another end-user of the provided health service) and a service provider laboratory create cryptocurrency wallets to be able to purchase and deliver services. Then, the diagram illustrates the case when a patient buys cryptocurrency to pay for a test. Such a test is auctioned among the registered labs, which bid automatically depending on the characteristics of the test. The criteria to select the auction winner may include the type of test, its cost, result accuracy or the service delivery time. Thus, labs can compete in an automated way to provide the most attractive services and the blockchain-based platform is eventually able to select the lab whose offer best suits the patient request.

Once the test lab is selected, the platform allocates the test to the winning lab and confirms the allocation to the patient/user, who pays for the service and indicates the shipping data. Then, the platform notifies the lab about the payment and the shipping details, so the lab can send a sample collection kit by post. Together with the sample collection kit, the lab provides a test ID and a password, which are used to guarantee private access to the TRs. The lab stores no data on the patient's real identity, just the wallet ID that performed the payment (which is an anonymous unique identifier), the test ID and the access password to the results. In this way, end-users can request health services in an anonymous manner.

In the next steps of the process the patient collects a sample, sends it by post and the lab processes it and obtains the results. Such results are stored in the decentralized storage (i.e., in a OrbitDB/IPFS node), protected by the password previously indicated to the patient/user. Then, the availability of the results is notified by the lab to the patient. Such a notification is performed through the blockchain-based platform to preserve anonymity: the lab carries out a transaction on the blockchain directly to the patient wallet ID indicating the test ID whose results are available. Then, the patient, who will be monitoring the platform periodically, will see a new notification on the availability of the results. In that moment, the patient/user will be able to access the results by accessing the decentralized storage system through a URL and the password previously provided by the lab. Finally, after the transaction finishes, the platform can provide another URL to the patient/user to provide feedback on the overall experience with the lab, so that other users can benefit from such experience in future interactions.

*3.5. Main subsystems and components*

In order to comply with the aforementioned requisites, the proposed blockchain-based DApp needs the following subsystems:

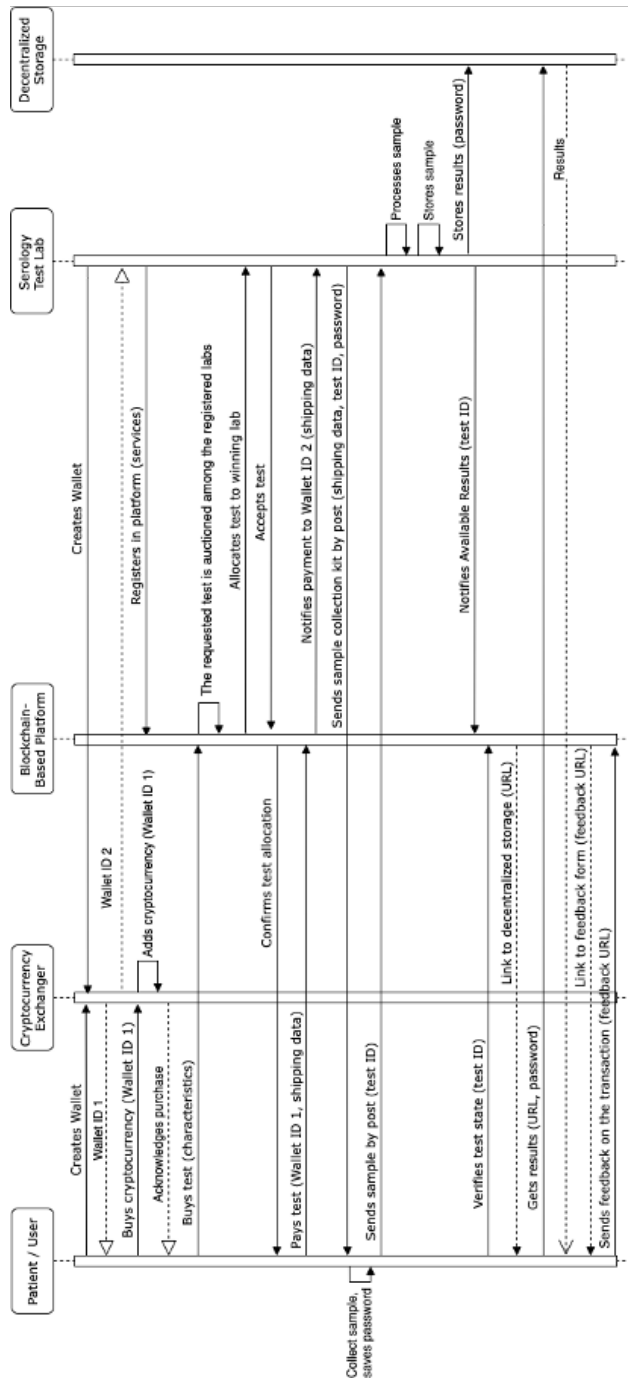

Figure 4: High-level sequence diagram of the blockchain-based lab test procedure.

- **Blockchain**: a blockchain network where smart contracts can be deployed.

- **Decentralized Storage**: blockchains are not meant to store large chunks of information, while centralized storage subsystems are impacted by the problems indicated in Section 2.3. As a consequence, technologies like those based on decentralized mechanisms like Inter-Planetary File System (IPFS) are able to provide a more secure alternative to store TRs.

- **Oracle**: this component checks the validity of off-chain data sources (e.g., external databases, real-time data), since blockchains are not able to make this verification process by themselves [58].

- **Privacy Layer**: IPFS does not automatically secure data, so it is up to developers to add encryption mechanisms to make sure that the DApp meets the security standards or legal requirements. In this case, the proposed health platform requires a way for sharing TRs between health providers and patients in a secure way.

The approach followed by the solution proposed in this article consisted in implementing the previously enumerated subsystems as independent Docker containers [59]. This allows an iterative approach, where the platform is first implemented and tested in a private local environment before its deployment in a real scenario (i.e., before using third-party providers or public blockchain networks). Moreover, such an approach allows any future researcher/developer to easily reproduce the entire system on his/her own computer.

Figure 5 depicts how the created Docker containers communicate. The following description only provides a high-level description of each component implementation, but the interested reader can find the software and a complete guide on how to deploy and integrate the containers at HELENE's official repository [15]:

- **Ethereum Private Network** [60]: an Ethereum-based private blockchain network component is used to harness the flexibility of Solidity smart contracts. The use of a Docker containerized testing environment provides more insight into blockchain processes compared to when using development endpoints [61]. Figure 5 also shows the contracts of the system, which are automatically synchronized across the Ethereum nodes. Such smart contracts define the logic or the back-end of the system:

- **Auction**: a test request is represented by the Auction contract, so there is one contract per request. Once the auction starts, the patient public address, the expiry date, the type and the preferences for the test request are established. Ending the auction will add the winning bid details and the public address of the winner laboratory to the test request. This contract allows for adding more complex logic, like the inclusion of an access control or the definition a specific bidding process.
- **Token**: this contract enables creating and using tokens, which are digital assets that act as real flat currency. The use of tokens allows for implementing incentive mechanisms to foster participation and data sharing. This contract is further explained in Section 5.1.
- **Oracle**: this contract is used to manage data exchanges between the DApp and the IPFS Network in a secure way. The use of this contract is illustrated later in Section 5.2.

- **IPFS Network** [62]: IPFS is a decentralized protocol that allows for building a fully distributed data storage, avoiding SPoF and ensuring data integrity [63]. To harness the benefits of IPFS, an IPFS-based local network component was created to automatically share and synchronize data among all the deployed instances. However, IPFS lacks built-in querying or filtering capabilities, and it has some limitations regarding data management (e.g., any kind of information can be stored in IPFS without format limitations). This makes it more appropriate for large amounts of data that are not updated frequently (e.g., TRs) and limits its use for a significant part of front-end data (e.g., user profiles, activity logs, events, ratings, reviews).

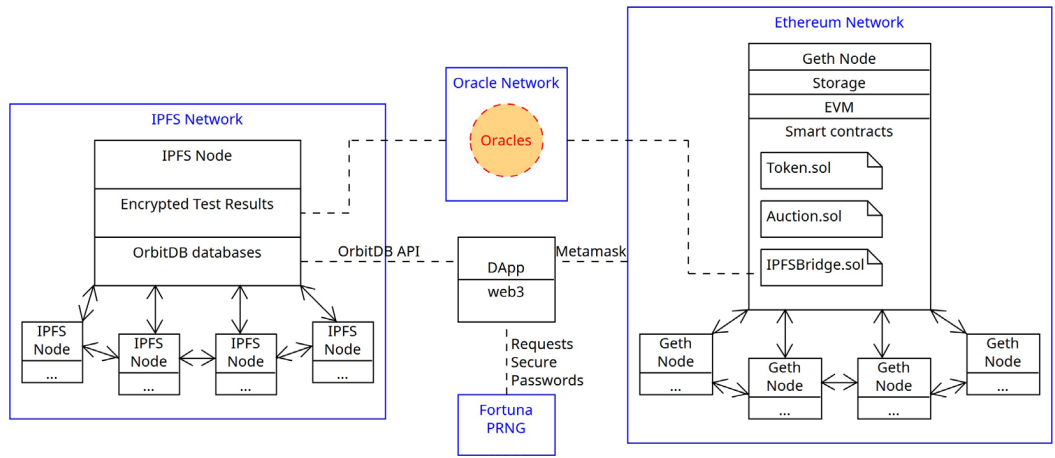

Figure 5: Component architecture of the system (each blue rectangle represents a Docker container).

- **Oracle** [65]: blockchains lack a native method to verify external data, so a custom oracle implementation was developed and containerized to establish a secure communication between the encrypted TRs stored on IPFS and the blockchain. This development is critical, since it allows for avoiding third-party oracle network providers and thus provides full control over the data interaction process.

- **Pseudo-Random Number Generator (PRNG)** [66, 67]: when a TR is uploaded, the system encrypts it with a password that allows for checking its information. Therefore, the password generation mechanism should ensure a high-degree of unpredictability to avoid potential attacks.

As both the Oracle and the PRNG components play a significant role in the proposed system, they are further detailed in Sections 5.2 and 5.3, respectively.

*3.6. Main Platform Processes*

*3.6.1. Patient Request Assignment*

Patient request assignment starts when a patient creates an auction for a test request with the desired type (e.g., PCR, Antigen or Antibody test), delivery time (e.g., expected time for delivering the test results) and accuracy (i.e., result accuracy percentage). This process will create a transaction in the blockchain that will deploy the Auction contract, automatically starting the auction.

Health providers (e.g., laboratories) can bid for that auction indicating their offered delivery time, price and accuracy. The different bids will be shown to the patient, who will be able to choose the best one according to his/her preferences. Note that this selection process can be easily automated (e.g., by choosing automatically the closest offer to the patient request), but since one of HELENE's aims is to create a patient-centered solution, the current version of the platform is implemented so that patients are actively involved in the final decision.

Once the offer is selected by the patient, the chosen lab will be notified that a patient is interested in its offer, so it has to confirm that it accepts the test request. After such an acknowledgement, the patient is notified and requested to pay the specified price with the platform token.

All the previous interaction is illustrated in the UML sequence diagram shown in Figure 6.

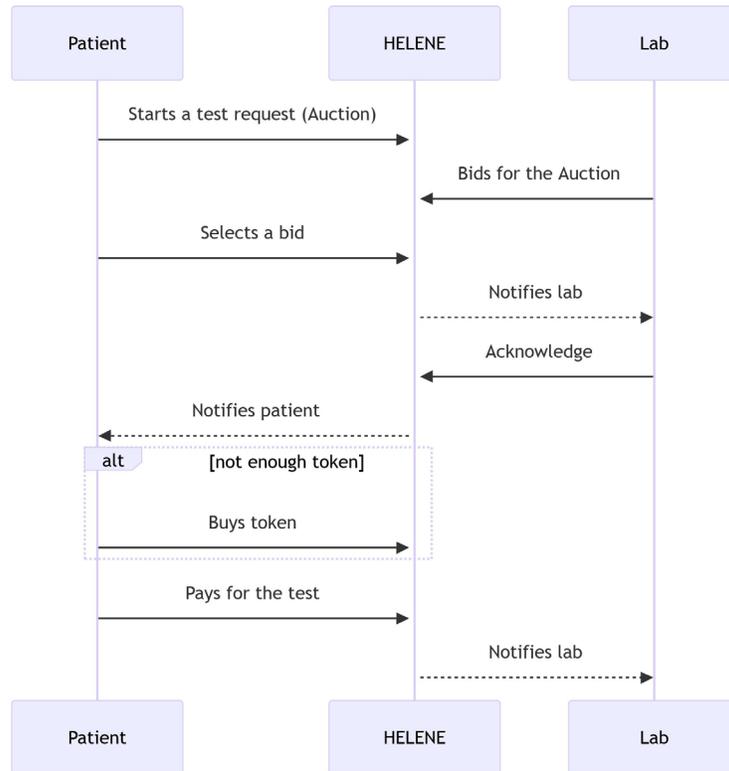

Figure 6: Patient bidding process.

*3.6.2. TR Management*

The TR management process is illustrated in Figure 7 and is as follows. Once an auction is allocated to a laboratory, it needs to send a sample col-lection kit to the patient. Then, the patient collects the sample and sends it back. Once the sample is processed by the laboratory, it uploads the TR, making it automatically available to the patient.

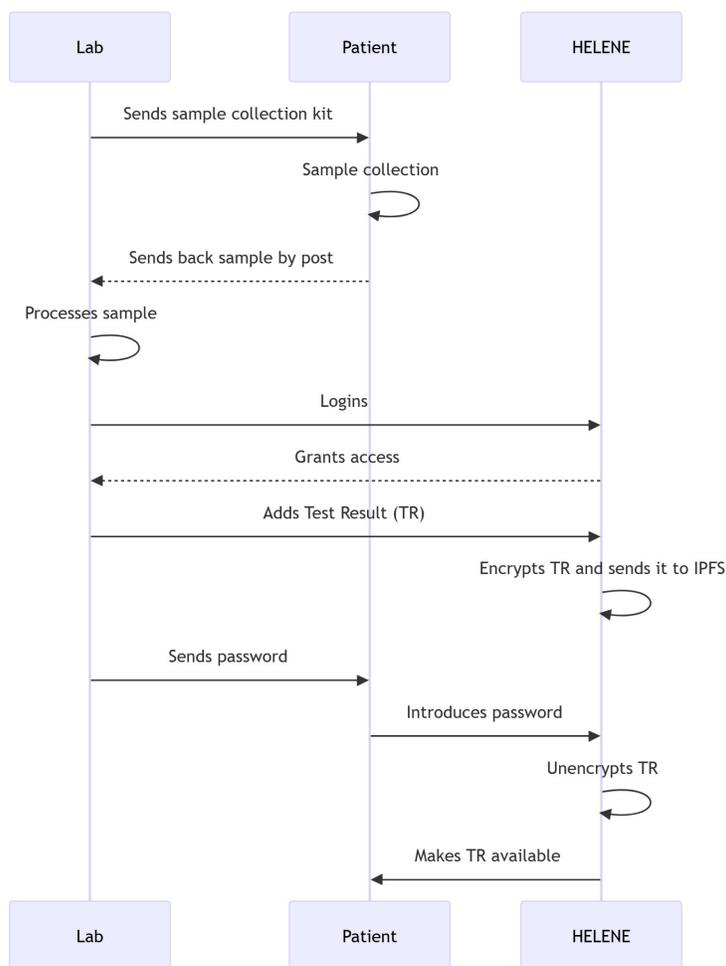

Figure 7: TR generation and delivery processes.

*3.6.3. Incentive mechanism*

Although Section 3.4 summarizes the main interactions of the blockchain-based platform, more complex functionality can be provided, like the automatic handling of private data (e.g., to control who has access to each patient health data) or the use of incentive mechanisms to reward patients that provide their data to, for instance, carry out research studies.

In fact, the proposed blockchain-based platform goes several steps beyond traditional digital health platforms and provides incentivization mechanisms to provide value to the patients/users and foster private health data sharing in a transparent and fair way.

An example of one of the way an incentivization mechanism would operate is illustrated in Figure 8. Such a Figure indicates the main steps involved in a situation when a pharmaceutical company wants to develop a product (e.g., a vaccine) and needs to acquire certain health information from a significant number of patients that have previously performed a specific test.

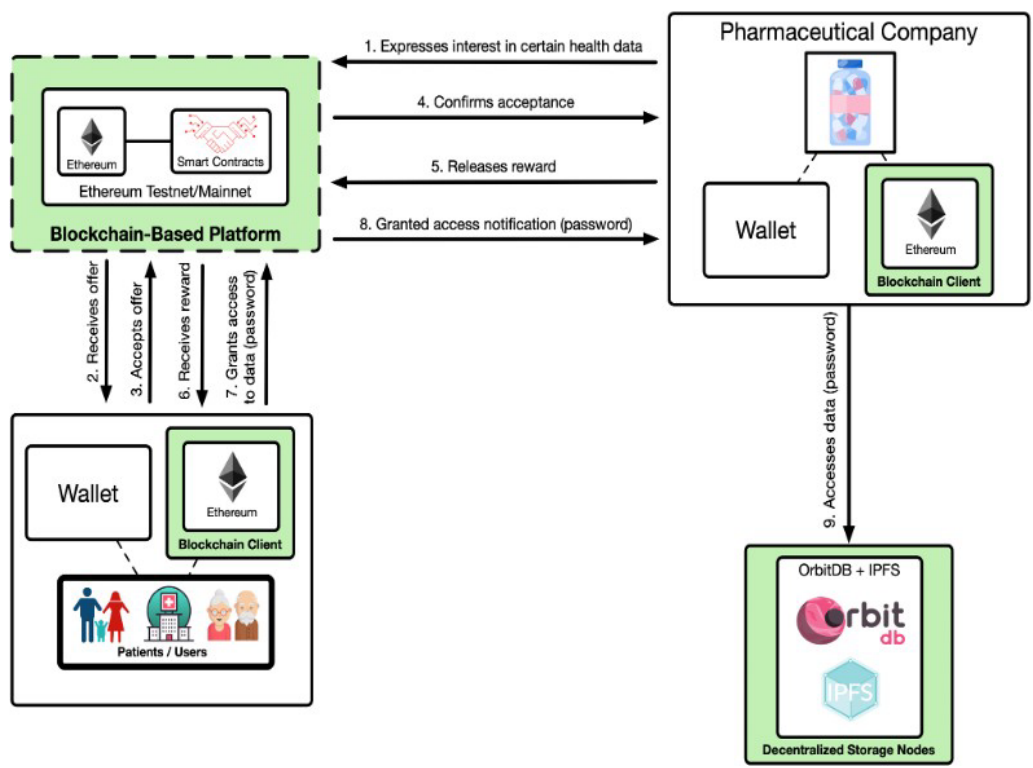

Figure 8: Example of HELENE's incentivization mechanism.

In such a scenario, the pharmaceutical company will express its interest on certain data to HELENE's blockchain-based platform (step 1 in Figure 8), indicating which reward (e.g., a specific amount of cryptocurrency) will be given to the data contributors. Such an offer will be redirected to potential patients/users that fulfill the data requirements of the company (2). Then, the users that accept the offer (3) will notify it to the company (4), which will send the agreed reward (5 and 6). Finally, the user will grant access to the required data, which will be protected with a password (7). After receiving the granted access notification (8), the company will be able to access the patient data using the provided password (9). These interactions are illustrated in the sequence diagram shown in Figure 9, where a lab pays with HELENE's tokens in exchange for patient data.

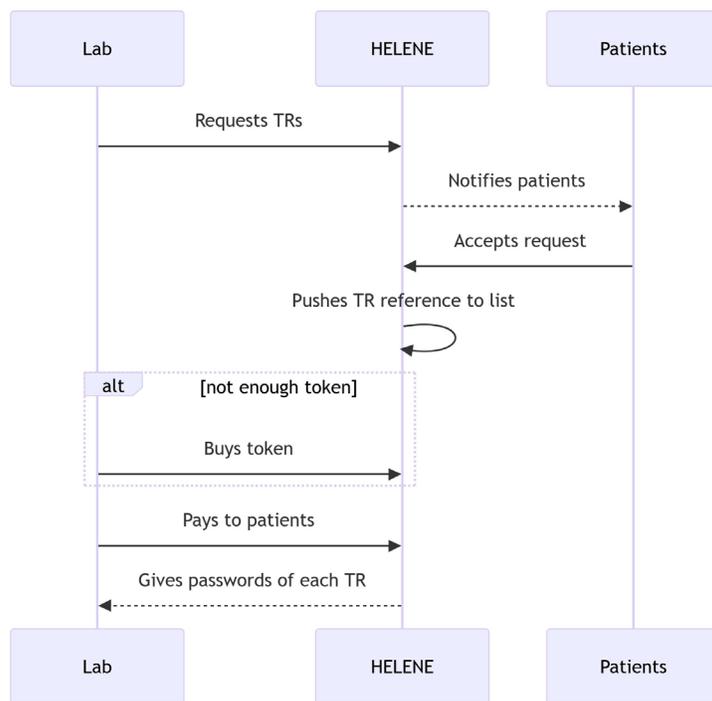

Figure 9: Patient token remuneration procedure.

## 4. Lab test procedure and test machine modeling

To demonstrate HELENE's capabilities, a novel lab test procedure aimed at detecting COVID (i.e., SARS-Cov-2) was selected. Such a procedure is described in the next subsections and consists of two parts: (1) the test procedure itself and (2) how the data obtained from multiple real tests were used to model them statistically so as to simulate the behavior of the test machine. It is important to note that the simulator was developed only for demonstration purposes and to evaluate the performance of HELENE, but it is straightforward to integrate the modeled test machine as if it was an Internet of Things (IoT) device.

*4.1. Test procedure*

*4.1.1. Blood sample collection*

The first step in the procedure is to collect blood samples from the patients. Such blood samples were collected in microcentrifuge tubes and allowed to coagulate undisturbed at room temperature for 30 minutes. The samples were centrifuged in a refrigerated centrifuge at 1,000-2,000×g for 10 minutes to remove clots. Following centrifugation, the serum was carefully pipetted into clean Eppendorf tubes. During handling, the serum was maintained at 4°C. For prolonged storage or transportation, the samples were kept at –20°C or colder, ensuring that multiple freeze-thaw cycles were avoided to preserve serum integrity. Samples showing hemolysis, icterus or lipemia were flagged as potentially compromising test accuracy and hence excluded from further analysis when necessary.

*4.1.2. Test Process*

To assess antibody responses, SARS-CoV-2 S1-coupled microbeads were aliquoted into 1.5 mL Eppendorf tubes (Eppendorf Cat#525-0133) with 50 μL of PBS, containing 5000–6000 beads per tube. After centrifugation at 2000 rpm for 5 minutes, the microbeads were resuspended in 100 μL of serum samples pre-diluted at 1:100 in PBS. Negative controls using PBS alone were included. Samples were incubated for 30 minutes at room temperature, protected from light, and washed three times with PBS. Secondary antibody labeling involved the addition of anti-human IgG-PE (1:50 dilution) (Clone G18-145, BD Biosciences Cat#555787) prepared in 100 μL of PBS with 5% FBS. Samples were incubated with this solution for 15 minutes at room temperature, shielded from light. Following a final wash, the microbeads were resuspended in 200 μL of PBS containing 5% FBS. Data acquisition was performed using an Attune NxT flow cytometer (Thermo Fisher Scientific), capturing a minimum of 600 events per microbead type. For each antigen the geometric mean fluorescence intensity (gMFI) was recorded and analyzed using FlowJo version 10 (BD Biosciences). The used equipment setup is shown in Figure 10.

The obtained gMFI values were classified as follows: values below 100 were considered negative, values between 100 and 200 were considered as inconclusive, and values exceeding 200 were categorized as positive.

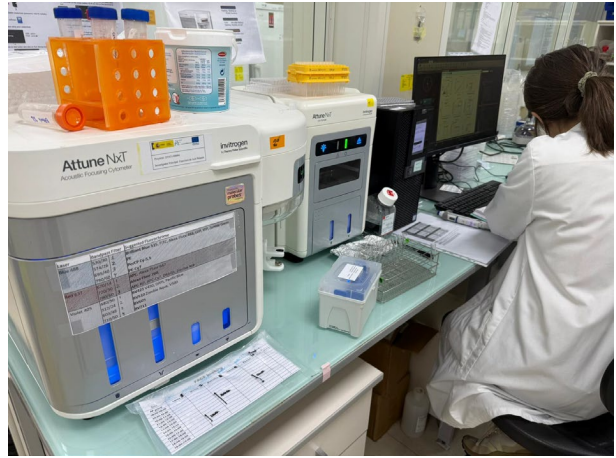

Figure 10: Equipment used for testing antibody response.

## 4.2. Test machine

Flow cytometry was performed using an Attune NxT Flow Cytometer equipped with 3 lasers (405 nm, 488 nm, 640 nm). This high-performance system features advanced sensitivity and resolution, enabling precise detection of fluorescence signals even in complex biological samples. The instrument capacity to process large sample volumes efficiently with a HTS autosampler makes it ideal for diagnostic applications, such as the quantification of SARS-CoV-2-specific antibodies.

## 4.3. Test machine modeling

The statistical modeling of the test machine was performed based on real data collected in 2020 and 2021 from 162 patients from the Spanish cities of Santander and Bilbao. By using such data, a distribution model was developed to realistically estimate the number of antibodies that a hypothetical test in a real environment could detect.

Figure 11 shows as an histogram the data distribution of the obtained antibody counts. The selected intervals were chosen on a simple inspection of the data distribution and considering the SARS-CoV-2 infection thresholds defined by the test procedure (i.e., $[0, 100]$ for non-infected patients, $[101, 200]$ for inconclusive cases, and $[201, \infty]$ for infected patients).

In the histogram it can be observed that there is the large number of patients without SARS-CoV-2 antibodies, which gradually decreases as the values approach the inconclusive region ([101, 200]). According to the dataset, there is a relatively low number of antibodies among infected patients. However, this trend reverses for values exceeding 1000 antibodies, where a significant and sharp increase is observed.

To make use of the obtained data distribution, it was converted into a probability density function. This allows for generating antibody counts with the statistical probability expected for the analyzed dataset and from the real test machine, thus enabling diagnostics to be made based on the previously described thresholds.

Once the diagnostic process is completed, the test machine generates the TR (a PDF file) that is delivered to the patient. At the same time, the TR is encrypted with the password generated by the PRNG (whose implementation is later described in Section 5.3) and stored on IPFS.

## 5. Implementation

### 5.1. Token Smart Contract

The standard chosen for the token implementation was the ERC-1400, since it provides more key features than the ERC-20 and more flexibility than the ERC-3643. However, ERC-3643 supports batch operations, a useful feature not included in ERC-1400 that is useful in terms of increasing efficiency. To address this limitation, HELENE's implementation of ERC-1400 was modified as follows to incorporate batch operations as well as other optimizations:

- Partitions represent disjoint subsets of tokens, governed by a set of rules defined during token implementation. This feature was originally conceived to simulate privileges associated with a set of stocks, such as voting rights in a corporation. This feature was excluded from HELENE's final implementation because it introduces unnecessary complexity, widens the compatibility gap with the ERC-20 standard, and fails to provide controller addresses with a direct method to withdraw funds.

- Batch operations were added to ERC-1400. This feature complements the standard transfer operations defined by the specification and does not detract from the generic transfer operations. Such an addition is based on the theoretical premise that processing a grouped set of transactions is more efficient on a blockchain than handling the same transactions individually on demand. Section 6.2 empirically validates this assumption with a performance comparison in terms of latency.

- A function was developed to facilitate the acquisition of tokens. Such a function is a transfer operation from an address that holds all available tokens. This address was selected because it initializes the smart contract on the network, giving it the authority to mint, to burn and to transfer tokens, as well as to delegate or to revoke these privileges to other addresses. This delegation is managed through additional functions designed to simplify privilege management.

- Finally, the implemented token also includes functions to potentially store document metadata on-chain and to provide rewards to laboratories and patients. It is important to note that, for the sake of brevity of this article, the analysis, design and evaluation of potential token-based incentivation strategies are outside the scope of this article.

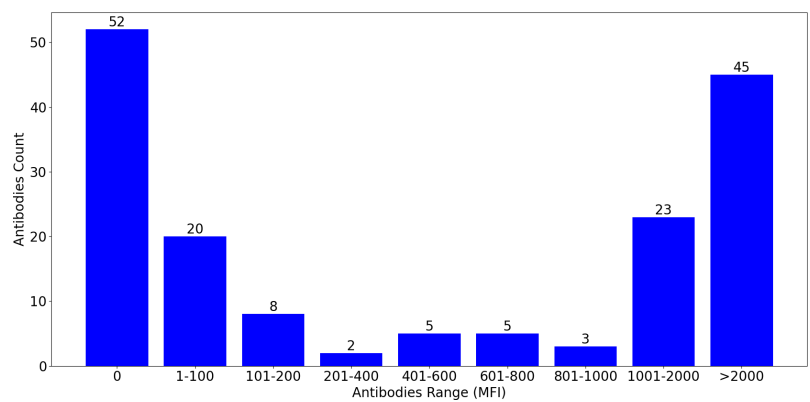

Figure 11: Histogram of the antibody count obtained by the test machine during the Santander and Bilbao campaign.

*5.2. Oracle*

*5.2.1. Overall Operation*

The developed oracle follows a request-response pattern [68] that is composed by two main parts:

- **Oracle Contract**: this contract is used to listen to requests from other smart contracts that need to access external data. Thus, the Oracle Contract acts as an intermediary, transmitting the received requests to external services that are deployed off-chain.

- **Oracle Nodes**: they are located off-chain and are responsible for processing requests received from the smart contracts executed on the blockchain. There can be multiple Oracle Nodes working in parallel to ensure redundancy and reliability, interacting with IPFS local networks to upload and to obtain the requested data. Such data are transmitted back from the Oracle Node to the Oracle Contract (i.e., to the blockchain), completing the request-response cycle.

The communication between the Oracle Contract (executed on the blockchain) and the external Oracle Nodes is carried out through events. When the Oracle Contract needs to interact with IPFS, it emits an event on the blockchain that contains all the information necessary for the interaction with IPFS. This information, once emitted, is stored permanently on the blockchain, allowing the Oracle Nodes to access it at any time.

Oracle Nodes are also responsible for the continuous monitoring of the Ethereum network where the Oracle Contract is deployed. To do this, the nodes subscribe to the contract events using its Application Binary Interface (ABI), thus being able to identify and to process the emitted events. In this way, when a node detects a specific event, it will process its arguments and information to perform the necessary interaction with IPFS.

*5.2.2. Implementation of the Oracle Smart Contract*

The main objective of the Oracle Smart Contract is to facilitate the interaction between the blockchain and the Oracle Node, as it is depicted in Figure 12. The following are its main components:

- **trigger IPFS Add Event**: this function allows the IPFS Add Event to be emitted, which sends a contract address and a cipher string to the blockchain. Such an encrypted value is the TR document in the form of a string cipher. The encryption uses AES-256 with the library CryptoJS [69]. The IPFS Add Event event is responsible for notifying the Oracle Node the storage of the specified encrypted value and the registration of the Ethereum account that will be authorized to retrieve such a value in the future. This event serves to start the process of storing encrypted data in the IPFS system, ensuring that both the cipher string and the sender of the request are recorded, which can be useful for auditing and access control.

- **trigger IPFS Cat Event**: this function is called whenever a user wants to access a specific TR. The contract then emits an IPFS_Cat_Event event, which tells the Oracle Node that it wants to retrieve the encrypted value associated with a specific address, allowing the oracle to begin the recovery process.

- **setValuesIPFS**: this function is used by the Oracle Node to store in the contract an encrypted address-value pair, which has been retrieved in response to the IPFS_Cat_Event event. Upon receiving a retrieval request, the oracle uses setValuesIPFS to record the encrypted value associated with the specified address on the blockchain.

- **getValuesIPFS**: this function allows users to retrieve the encrypted value associated with a specific address, once it has been stored in the contract by using the setValuesIPFS function. getValuesIPFS is a read-only function that returns the encrypted value stored in the contract for a given address, thus facilitating the querying of previously recorded information. Note that this function can only be invoked after the trigger_IPFS_Cat_Event sets the corresponding value.

*5.2.3. Implementation of the Oracle Node*

To implement the previous functions of the Oracle Node, Python was used along with the Web3.py library, which facilitates connecting to the Ethereum network and subscribing to Oracle Contract events. Using both the address of the contract deployed on the Ethereum network and the contract ABI, the Python code is able to understand the structure and methods of the smart contract, allowing for precise interactions. Events (such as IPFS_Add_Event and IPFS_Cat_Event) and functions (such as setValuesIPFS and getValuesIPFS) are defined in the ABI, allowing the node to communicate and manipulate data within the contract efficiently. This allows the Oracle Node to monitor emitted events in real-time and to respond according to the parameters of the detected event. Thus, the tasks of the implemented Oracle Node can be divided into two groups: one with the tasks dedicated to initializing the IPFS node and another one with tasks focused on monitoring the blockchain network.

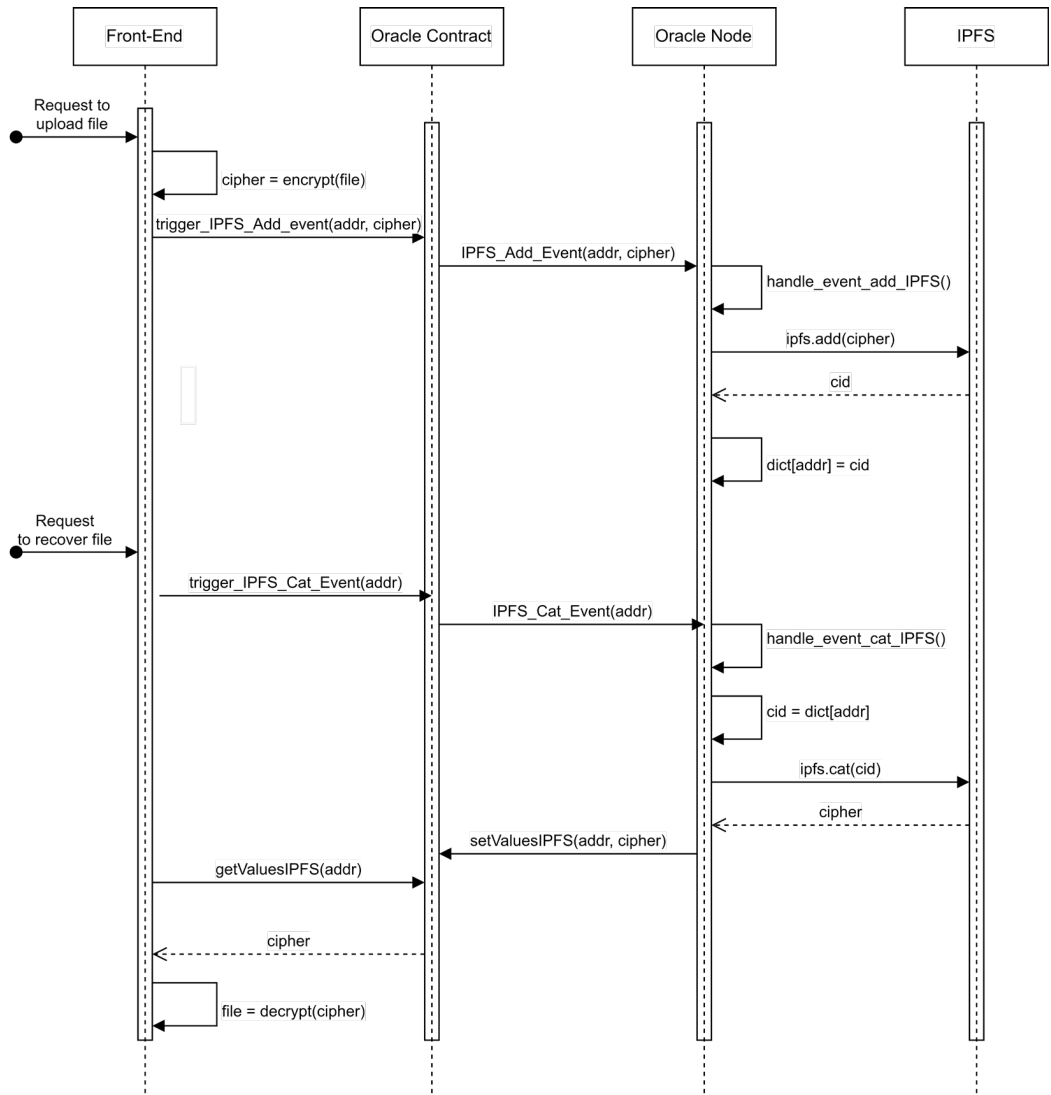

Figure 12: TR storage and retrieval process with the help of the Oracle.

To facilitate the deployment of the Oracle Node, a Docker container is used to run the IPFS daemon automatically, initializing the IPFS service in an automated way. Once the IPFS node is up and running, the Python program is executed and proceeds to connect to the Ethereum network using Web3.py and the contract details, including its contract address and ABI. Thus, it will create an instance of the contract that will be used to interact with the blockchain. Thanks to the web3.py library, a series of filters can be created to allow for the detection of emitted events. Once the previous steps are completed, the Oracle Node is ready to monitor events on the network. When an event is detected, the corresponding functions will be executed based on the type of emitted event.

It is worth mentioning that the oracle can be deployed as a single IPFS instance, thus simplifying the integration process and providing a controlled environment for testing, iteration and optimization. However, for real deployments, a decentralized oracle is more secure and can be also efficient, as it is later analyzed in Section 6.3.

### 5.3. Secure PRNG API

#### 5.3.1. Main components

Typically, encryption mechanisms involve the use of passwords, which provide controlled access to files. In the case of TRs, they can be encrypted so the access is granted only to users with the correct password. The generation and use of such passwords must effectively address the problem of real randomness and be resistant to brute-force attacks.

Passwords should be generated from pseudo-random numbers, avoiding problems commonly associated with human-created passwords, such as password reuse, adding personal information (e.g., names, birth dates or common words) or using sequential patterns.

Since blockchains are inherently deterministic, it is not possible to generate random numbers directly within the blockchain [70]. Entropy sources in Ethereum are limited and open to potential manipulation, as common methods for generating randomness, such as block hashing or block timestamp, can be exploited or manipulated by attackers to obtain predictable results [71]. Moreover, this approach would also increase the cost of executing a smart contract, which is an important factor to consider in terms of efficiency. Furthermore, due to the public nature of the blockchain and the importance of security in generating passwords, using an oracle to generate and to transmit the random generated numbers is also impractical, since they would be exposed on the network, putting its security at risk. For this reason, the randomly generated numbers must be generated and transmitted externally to the blockchain. This goes against the full decentralization fostered by certain blockchains, but it allows for providing a more robust entropy generation method that is less susceptible to manipulation.

*5.3.2. Secure password generation*

For secure password communication, the solution presented in this article makes use of a REST API protected by a Zero-Knowledge Proof (ZKP). In such a system, the user signs a randomly generated message with MetaMask [72]. Based on this signed message, the authenticity of the signature is verified, confirming whether the account that has signed corresponds to one of the authorized accounts. By using digital signatures as an authentication method, a high level of security is guaranteed, as the private keys required to sign messages are protected and never shared, significantly reducing the risk of account compromise.

When it comes to the length and characters used in generating the passwords, the balance between security and ease of use has been carefully considered. Password length is one of the key factors influencing its strength, as a greater number of characters exponentially increases the possible combinations, thus making brute force attacks more difficult. The character set also plays a key role in password security. If the set of characters specified in Table 2 is used, a total of 70 possible characters are available, which can be divided into four groups: lowercase letters (26 characters), uppercase letters (26 characters), numbers (10 characters) and special characters (8 characters), each contributing to the diversity and complexity of the character set.

| Characters type | Example | Character set |
|---|---|---|
| Lowercase | a-z | 26 |
| Uppercase | A-Z | 26 |
| Numbers | 0-9 | 10 |
| Special characters | !, ?, &, @, *, %, $, # | 8 |

Table 2: Password character sets.

To calculate the total number of possible combinations, the following equation was used:

$$s = n^l \tag{1}$$

where $s$ represents the total number of combinations, $n$ is the total number of characters in the set, and $l$ is the length of the generated password.

While a minimum length of 12 characters is recommended for a secure password, as long as they are chosen randomly from a sufficiently large set [73], this length may eventually fall short due to the increasing progress in processing power. Due to this fact, it is adequate to increase password length to ensure resilience against future brute-force attacks.

In this context, using the defined character set and a length of 16 characters, the total number of possible combinations is calculated as $70^{16}$, yielding an extremely large number. This number of combinations dramatically increases security against brute-force attacks and ensures high entropy, making the password significantly more challenging to crack, even with advanced processing technologies.

*5.3.3. Implemented PRNG: Fortuna*

When choosing a PRNG, it is important to guarantee security, efficiency and speed, so that this component does not affect or slow down the correct functioning of the overall system. Among the potential algorithms to be used, Fortuna [74] was selected, which is a cryptographically secure random number generator designed to provide a source of high-quality random numbers, and which was created to replace the Yarrow PRNG [75].

Fortuna is composed of three main elements:

- **Entropy accumulator**: it collects random data from various sources and uses it to restart the generator when an adequate level of randomness has been reached. The design of Fortuna allows it to take advantage of different sources of entropy.

- **Generator**: it takes a fixed-size seed and produces an indefinite amount of random data. On each request, it generates 256 bits of additional random data, which are then used as the new key.

- **Seed file**: an independent file filled with entropy ensures that the PRNG continues to generate random data even after being restarted. This file is used to initialize the generator.

Eventually, the pseudo-random numbers generated with Fortuna are converted into a series of valid characters for passwords using the character set defined in Table 2.

*5.3.4. Fortuna implementation*

The Fortuna algorithm was implemented from scratch in Python, primarily due to two reasons: (1) to ensure the algorithm operation closely follows the design proposed originally by its creators and (2) due to the lack of Fortuna implementations in the most widely used cryptographic libraries. The implementation focused on two key components of the algorithm: the Generator and Accumulator classes. However, the implementation of the seed file was not included. Such a seed file is intended for system restarts but, since the solution proposed in this article is designed to operate continuously (with rare restarts), the seed file was considered unnecessary. Moreover, eliminating the use of the seed file significantly reduces the constant write operations to this file that occur every time pseudo-random numbers are generated. Therefore, by eliminating the seed file system, performance is increased, achieving much more efficient pseudo-random number generation and decreasing resource usage.

*5.3.5. ZKP implementation*

As it was previously mentioned, the authentication uses a ZKP to secure the API responsible for managing the secure communication of the system passwords. Such a method enables user identity verification without the need for sharing sensitive information, such as his/her private key. In order to authenticate, the user will sign a message generated pseudo-randomly. This generation uses the system source of randomness, optimizing performance with each login request to the API. This approach ensures that even if an attacker intercepts the signed message, it cannot be reused, since the system generates a new message for each request, thereby minimizing the risk of message reuse.

The logic behind the message requirements follows the same principles as those described previously for the password generation policy. However, in this case, the length of the message is not limited, as the impact on performance is minimal due to the use of the system randomness source. Additionally, the message is self-generated and for one-time use, so the user does not need to enter or to memorize it. For this reason, instead of using 16 characters, the message makes use of 32 characters, which exponentially increase the number of possible combinations to $70^{32}$.

Each generated message is associated with a UUID (Universally Unique Identifier), an identifier widely used in computing systems to distinguish each message uniquely. UUIDs are generated using algorithms that ensure a high degree of randomness, making them difficult to predict and free of predefined patterns. This property of UUIDs enhances system security by associating each message with a unique identifier, further protecting against impersonation attempts or signed message reuse.

For this subsystem, Python was used to implement a REST API responsible for handling ZKP authentication, which allowed to integrate it easily with the Fortuna random number generator.

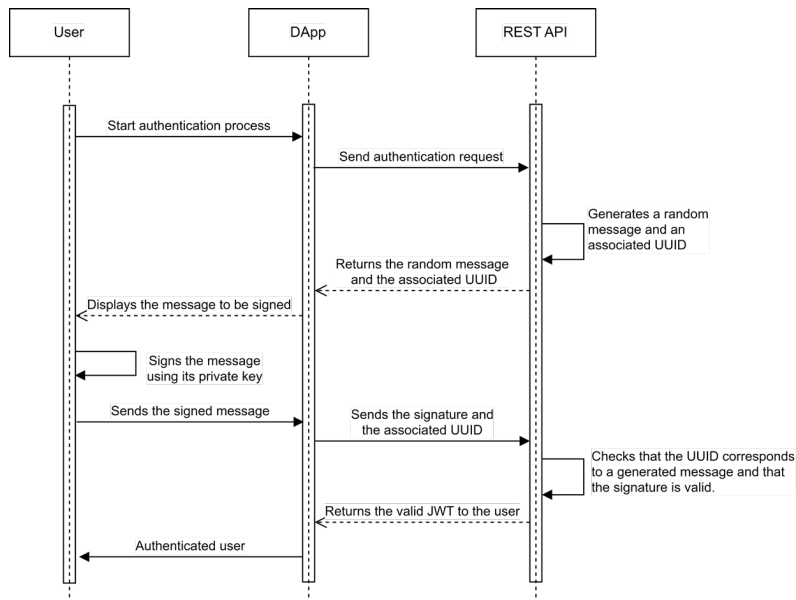

Figure 13: REST API authentication sequence diagram.

### 5.3.6. *Authentication*

The authentication process, illustrated in Figure 13, is as follows. First, the user initiates the process by clicking the authentication button, after which the application sends a request to the REST API to begin the authentication process. Then the API processes the request, generates a random message and an associated UUID, stores them, and returns them to the application as a response. The application receives the random message and displays it to the user, who signs it with his/her private key. The application then sends the signature and the UUID to the API, which verifies that the UUID corresponds to the one stored, retrieves the associated message, and checks that the signature corresponds to one of the public keys authorized to access the API. If this verification is successful, the API returns a JSON Web Token (JWT) to the application that allows access to the protected endpoints.

In this system, the JWT serves as a credential that clients use to prove that the authentication process was completed successfully. Once issued by the API, the token can be included in the next requests to access protected endpoints, eliminating the need for repetitive authentications.

Once a user receives a JWT, he/she will have access to the secure endpoint that will allow for requesting the API to generate passwords. Upon making such a request, the API will trigger the Fortuna PRNG, which will be responsible for producing a sequence of random numbers. These numbers will then be converted into characters that belong to the character sets indicated in Table 2. After generating the initial sequence, the API will perform a final check to ensure that the generated password meets all the security requirements indicated in the password policy. This check will ensure that the password has a minimum length, contains an appropriate combination of characters, and does not present predictable patterns. Once its compliance with the security policy is confirmed, the final password is sent to the user who made the request. This whole process is illustrated in the sequence diagram shown in Figure 14.

*5.4. Lab test machine simulator*

The laboratory test machine simulator was developed using an object-oriented approach in Python. It is composed by two main classes:

- **Evaluator**: it simulates the test procedure by generating a random number of antibodies and a diagnostic in accordance with the probability density function described in Section 4.3. The generated diagnostic, along with the number of antibodies, can be exported as a simple PDF report, which is then delivered to the patient via the devised decentralized mechanisms.

- **AnalysisOfferor**: it manages all networking capabilities, including querying data about available auctions, placing bids and submitting PDF reports.

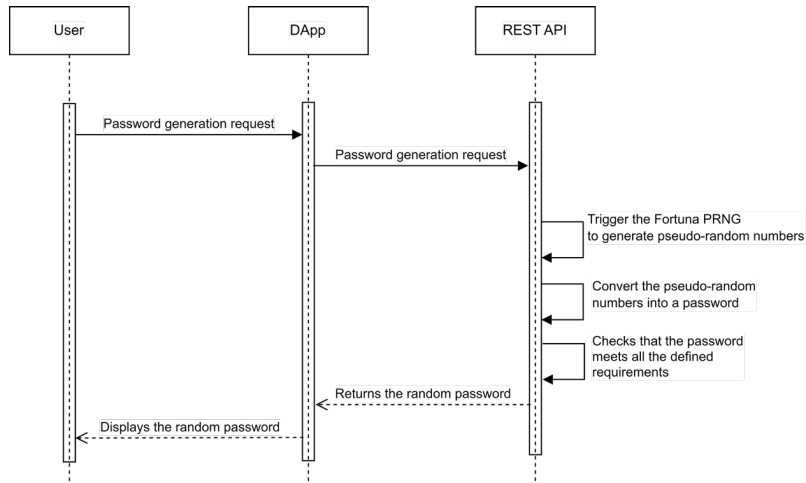

Figure 14: Password generation sequence diagram.

It is worth pointing out that the developed simulator is able to simulate the behavior of actual bidding labs, which is useful for carrying out tests or for recreating certain situations. Specifically, the simulator is able to get the available auctions and to bid for any of them. Moreover, it can send the TRs to the patients as PDF reports, which are signed digitally by the laboratory using its private key certificate, thus enabling the patient to verify the document authenticity off-chain. After signing, the report is encrypted with an AES symmetric key generated with Fortuna.

## 5.5. Front-end

### 5.5.1. Main features

The DApp front-end presents an interface layer that integrates usability, robust security and efficiency. Specifically, it is built with Next.js, a state-of-the-art React-based framework. Its design was conceived to ensure performance, responsiveness and adaptability across various device types. The main dashboard (shown in Figure 15) acts as a central hub, providing users with a comprehensive overview of the platform features and their current activities. It incorporates multiple functional panels, such as an auction interface that displays real-time data from the laboratory simulator. This enables users to track active auctions and bid statuses in a user-friendly format. Regarding the TRs menu, it is designed to streamline laboratory workflows, enabling secure upload and management of TRs. Thanks to this menu laboratories can interact with an easy-to-use upload interface, which transmits data securely to the back-end for processing. This menu also integrates the robust password generation utility powered by the Fortuna PRNG, ensuring the creation of cryptographically secure keys for encrypting PDF reports. Moreover, a wallet address display is dynamically updated to reflect the user's connected account, ensuring real-time synchronization whenever the account or network changes.

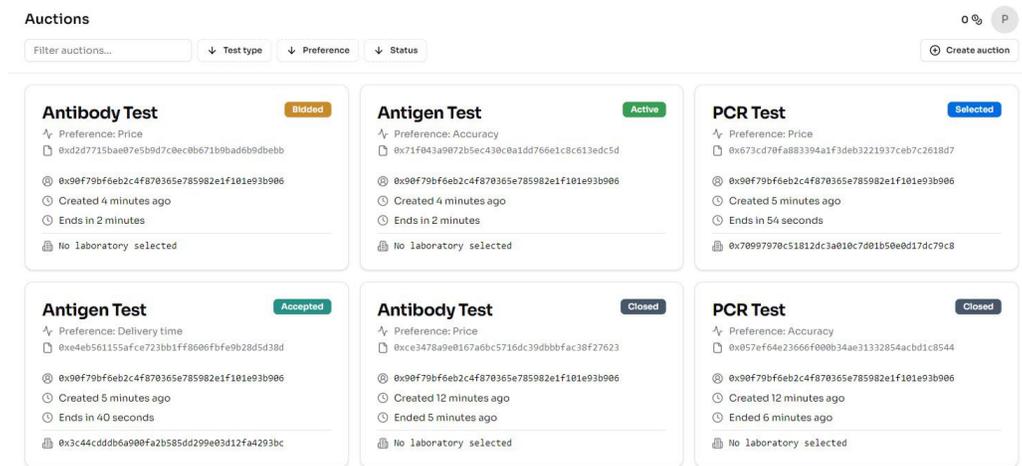

Figure 15: Real-time auction interface displaying active auctions and statuses.

*5.5.2. Example of interaction flow*

Figure 16 illustrates how a patient interacts with the developed platform since when a test is requested by such a patient and until the result report is shared with him/her. Specifically, the involved main steps are the following:

- Initially (step 1 in Figure 16), the patient accesses the platform and submits a test request, specifying the type of analysis required, along with their preferences regarding accuracy, delivery time and price.

- The patient request then enters the auction system, where laboratories submit their bids based on these criteria (step 2).

- Once the patient selects a laboratory (step 3), the system finalizes the assignment, and the laboratory receives the request details. Before

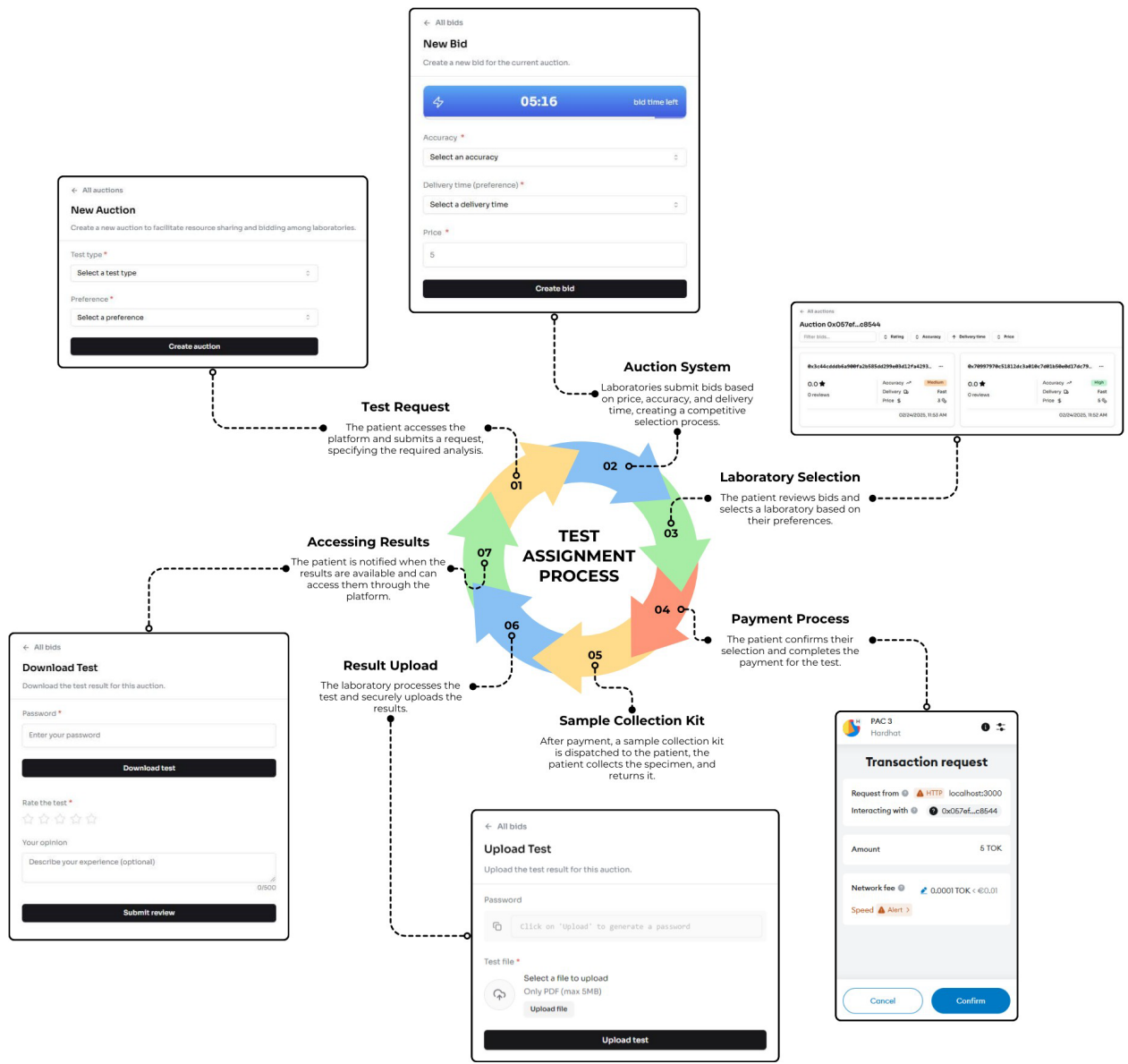

Figure 16: Interaction flow of the test assignment process in the HELENE platform.

proceeding, the patient makes a payment to the laboratory to confirm the transaction (step 4).

- After confirmation, the laboratory dispatches a sample collection kit to the patient, who collects the required specimen and sends it back (step

5).

- Upon receiving the sample, the laboratory processes the test and securely uploads the results (step 6). This upload mechanism ensures the integrity and confidentiality of the transmitted TRs, aligning with the system privacy-preserving principles.

- Once the test report is available, the patient is notified and can access the results through the platform (step 7).

## 6. Experiments

*6.1. System Performance*

To evaluate the performance of the system described in this article, the Holesky testnet [76] was used, since it effectively replicates the behaviour of the Ethereum mainnet using the Proof-of-Stake consensus protocol. The use of such a protocol means that the DApps deployed in the Ethereum mainnet are 'greener' than those deployed in past Proof-of-Work based networks [77].

It must be noted that, even though the involvement of a large number of nodes in a public blockchain enhances security, they impact transaction processing due to the generated communications overhead and block size. This trade-off implies a thorough study to decide which parameter is more important (i.e., decentralization, security or scalability) in order to choose the most suitable network for the DApp. Therefore, testing the solution in a mainnet "simulation" like Holesky can be seen as a worst-case scenario in terms of performance, before deploying the DApp contracts in a production network.

To carry out this set of experiments a simple script [78] was created to record the time from which a function is called until it is registered in a block. The tested operation was the deployment time of the contract, which is the most typical operation in the DApp. The results are depicted in Figure 17, where each recorded transaction duration is plotted as a point in a scatter plot, so the reader can see the variability of each emission. As it can be observed, the unpredictability of the behavior of the network leads to differences of up to 40 seconds for the same operation in some cases. Despite such outliers, the average latency remains at 13.27 seconds, which is an acceptable response time for the proposed application.

If the security of the platform is enhanced in the future by introducing the logic of the bidding system into a smart contract, further actions are necessary to avoid delays on bid registrations. In this regard is important to note that Solidity allows for starting an auction during deployment, which can potentially reduce the necessary number of transactions to the network. Moreover, it is worth noting that alternative EVM-based networks with efficient consensus algorithms can be used to tackle this lack of predictability while ensuring a low transaction latency [79, 80].

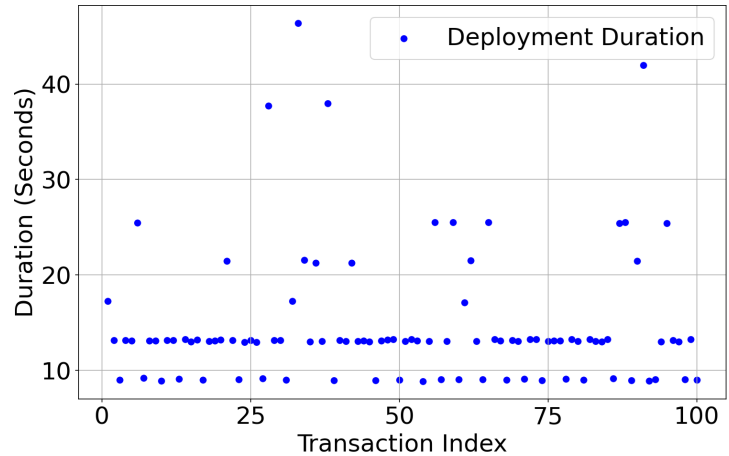

Figure 17: Deployment time for 100 sequential deployments.

*6.2. Smart Contract Optimization*

As it was described in Section 5.1, an ERC-3643 method was added to the Token contract to perform simultaneous transfers within a single transaction in what is called a batch transfer. Certain processes in the proposed health platform can be cumbersome for users without such an automation. For example, a batch transfer comes in handy when patients need to approve the external use of their data to several health providers, or when a laboratory wants to pay to several patients in exchange of data.

In this case, the performance comparison was carried out by using isolated endpoints (i.e., Hardhat) to avoid the external interference of public networks (e.g., network overhead, connections, response delays). Thus, the duration of sending 100 single token transfers was compared with the duration of sending a unique batch transfer that includes those 100 token transfers. It should be noted that, while performing 100 token transfers requires to sequentially send 100 individual transactions to the blockchain with the classical transfer operation, a batch transfer only requires to send one transaction. The obtained results are shown in Figure 18. As it can be observed, the communications overhead of the network can be effectively alleviated if less transactions are emitted. However, note that this operation is only valid for token transfers, not for other kind of processes.

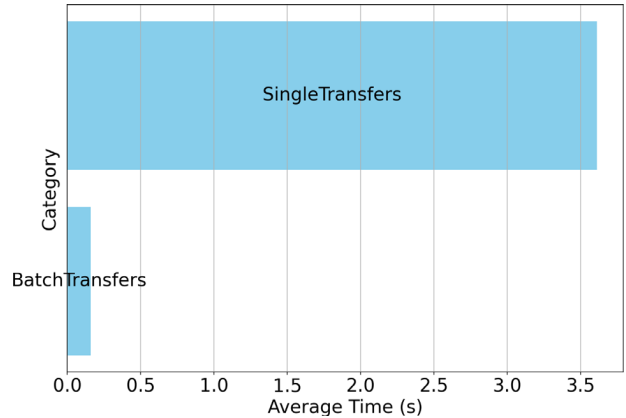

Figure 18: Average batch transfer time versus single transfer time for 100 token transfers.

### 6.3. Decentralized oracle performance

#### 6.3.1. Motivation

In the proposed system, the number of deployed oracles does not directly affect the file uploading process on the IPFS network, as the design guarantees operability regardless of such a number. However, there are other key aspects that are significantly influenced by the number of deployed oracles, especially with regard to system performance, security and availability.

On the one hand, a higher number of oracles strengthens the security and availability of the system, due to the redundancy provided by decentralization. This approach reduces the probability of the system being compromised by attacks targeting individual oracles or by failures in one or more of them. Redundancy ensures that even if some oracles are unavailable or malfunction, the others can continue to operate and ensure service continuity.

In addition, the decentralization of oracles allows for implementing a more sophisticated consensus mechanism, which is not based solely on the simple majority of results provided by the nodes. Instead, a model could be considered that includes additional factors, such as the reputation and history of each node. This approach would assign a different weight to the results provided by each node based on its previous performance, evaluating aspects such as its accuracy, consistency, and reliability in previous operations. In this way, a merit-based consensus system can be built, which prioritizes the contributions of the most reliable nodes. This not only increases the robustness of the system against attacks by malicious nodes or errors, but also incentivizes participants to maintain high quality standards in their behavior within the network.

On the other hand, the increase in the number of oracles has a direct impact on the performance of the network, both in the used blockchain and in the IPFS network. This is because a greater number of oracles generates more transactions, which increases the load on both networks:

- On the blockchain network: each oracle must record the transactions required to upload information, which can saturate the blockchain processing capacity, increase confirmation times and increase transaction costs.

- On the IPFS network: the additional traffic is also reflected in the storage network, as more oracles may involve simultaneous duplication or updating of data. This increases bandwidth and storage usage, which could affect the responsiveness of IPFS nodes.

Due to the previous issues, it is essential to find an appropriate balance in the number of used oracles, depending on the specific objectives of the system. For instance, a large number of oracles is ideal in scenarios where security and availability are paramount, such as in critical applications or where sensitive information is handled. However, in scenarios where speed and cost are important, it might be more efficient to limit the number of oracles to avoid overloading the involved networks.

Therefore, to make an informed decision on the number of oracles, it is important to consider factors such as:

- Blockchain network capacity: it is necessary to evaluate the level of scalability and transaction per second limits that the selected blockchain can handle.

- IPFS network efficiency: it must be considered the impact of data replication across nodes and the cost associated with distributed storage.

- Operating costs: a larger number of oracles implies more expenses associated with maintaining, synchronizing and monitoring them.

- Desired level of decentralization: depending on the degree of trust needed, the system may require more or fewer independent nodes operating as oracles.

- Risk tolerance: it should be evaluated how many oracles can fail without compromising system operation.

Therefore, choosing the optimal number of oracles should strike a balance between improved security, availability and the potential impact on system performance. This decision will depend on the specific use case and system priorities, considering factors such as the desired level of decentralization, available resources and the scalability of the blockchain network and IPFS involved.

*6.3.2. Experimental setup*

As it was indicated in the previous subsection, there are several ways to decentralize an oracle, each with advantages and limitations depending on the needs of the system. For the performed tests, it was decided to implement a simple but effective approach that introduces a basic level of decentralization and consensus between multiple nodes.

Specifically, the Oracle Contract was modified to store the responses of different nodes. Those nodes worked independently, individually uploading the same data to the IPFS network and reporting their results to the contract. Once the responses were received, the contract applied a consensus mechanism based on a simple majority. Such a mechanism evaluates the data sent by the nodes and selects the response that has been reported by the majority as the valid result. Eventually, once the valid result is determined, the contract records it and makes it accessible to the user.

Tests were performed to compare the response time between a centralized oracle and decentralized oracles. For each request, the same file was uploaded, so that the file size was not a determining parameter that affected the results. In total, 300 requests were carried out for each of the following cases:

- Centralized oracle with 1 node.

- Decentralized oracle with 2 nodes.

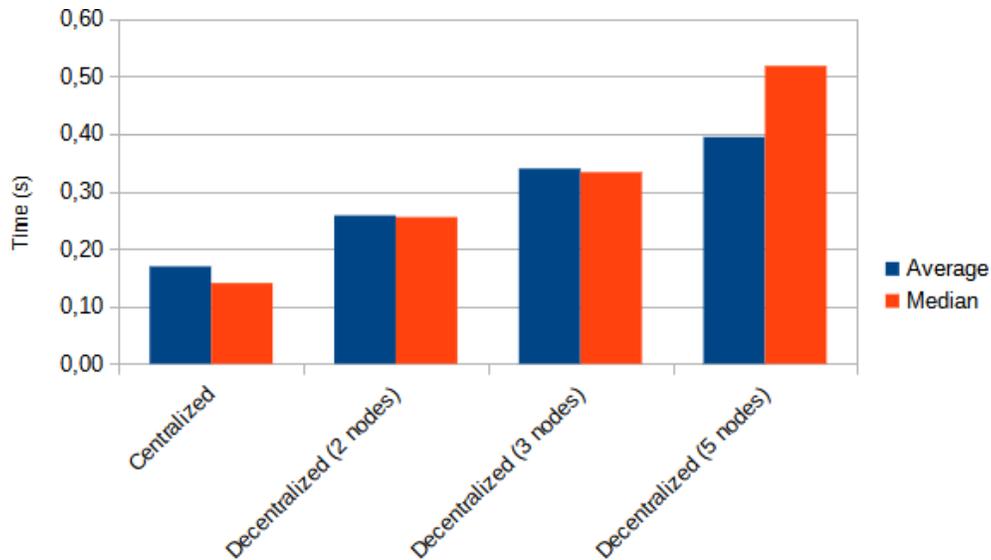

Figure 19: Average and median oracle response times.

- Decentralized oracle with 3 nodes.
- Decentralized oracle with 5 nodes.

For each configuration, it was measured the time elapsed from the upload request to the oracle until the event notifying the user that the file has been successfully uploaded and is available for access is generated.

*6.3.3. Number of decentralized nodes*

The obtained results are shown in Figure 19. As it can be observed, the centralized oracle has the lowest response times, both in terms of average and median. This is because in a centralized system there is no need to coordinate multiple nodes, which reduces latency and simplifies operation. By adding a second node, response times increase slightly. This is due to the need for synchronization between nodes and the additional processing required to ensure consistency in the system. With three nodes, response times continue to increase. For five nodes, response times are the largest compared to the other configurations. Therefore, the higher the number of decentralized nodes, the higher the provided security, but the longer the time required by the oracle to respond.

*6.3.4. Impact of the file size*

Tests were also performed to compare the impact of the file size uploaded to the IPFS network in relation to the response time of the nodes. These tests were carried out using two JSON files with test data: one of 12 KB and another one of 24 KB, to which AES-256 encryption was applied.

The obtained results are shown in Figures 20 and 21. According to the obtained data, as file sizes increase, response times grow significantly, especially in scenarios with decentralized nodes.

For the 12 KB file, when moving from a centralized to a decentralized network with 2 nodes, the average response time increases from 2.75 s to 6.89 s, which represents an increase of 150%. If the number of nodes is increased to 3, the average time reaches 12.03 s, which is equivalent to an increase of 337% compared to the centralized network.

Regarding the 24 KB file, when moving from a centralized to a decentralized network with 2 nodes, the average time increases from 6.11 s to 29.89 s, which represents an increase of 389%. If it is increased to 3 nodes, the average time reaches 49.21 s, with an increase of 705% compared to the centralized network. This shows that the impact of the number of nodes is more pronounced as file size increases due to the higher overhead in communications and synchronization.

Therefore, to optimize the network performance, it is critical to balance the number of nodes based on the size of the files. For small files, more nodes can be used without a severe impact on latency, improving the resilience of the system. However, for larger files, it is advisable to limit the number of nodes to reduce communication overhead, ensuring a better balance between response time and decentralization.

It should be noted that all tests were run on the same computer. Therefore, response times could be improved if a dedicated computer was used for each oracle. In that case, the improvement in the response time would be proportional to the capabilities of the hardware running each oracle.

*6.4. Random number generator NIST tests*

*6.4.1. Experimental setup*

To thoroughly assess the quality of the developed PRNG, the statistical tests specified by NIST SP 800-22 [81] were implemented and executed. Such a standard contains a widely recognized set of tests for measuring the randomness and security of pseudo-random number generators in critical applications. Specifically, the NIST SP 800-22 tests allow developers and researchers to analyze whether a sequence of numbers generated by ta PRNG exhibits the statistical properties expected from a truly random sequence. These properties include uniformity, independence and the absence of repetitive patterns. Thus, by subjecting the developed PRNG to the NIST tests, it is possible to identify whether there are unusual patterns or structures in the generated sequences, which could indicate weaknesses or vulnerabilities in the PRNG. This is critical, since any anomaly in the sequence could compromise the security of the system that depends on these numbers.

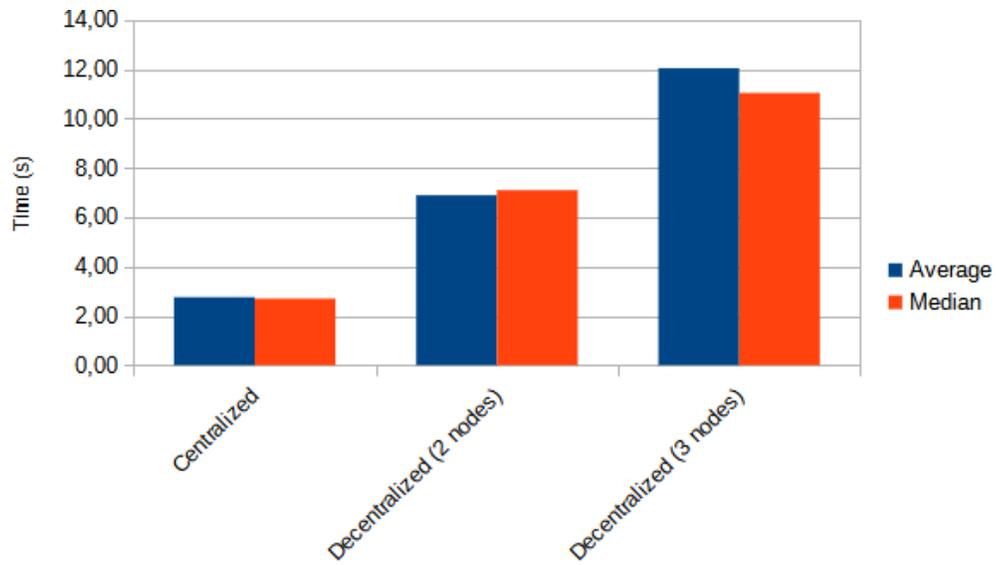

Figure 20: Response times for 12 KB files.

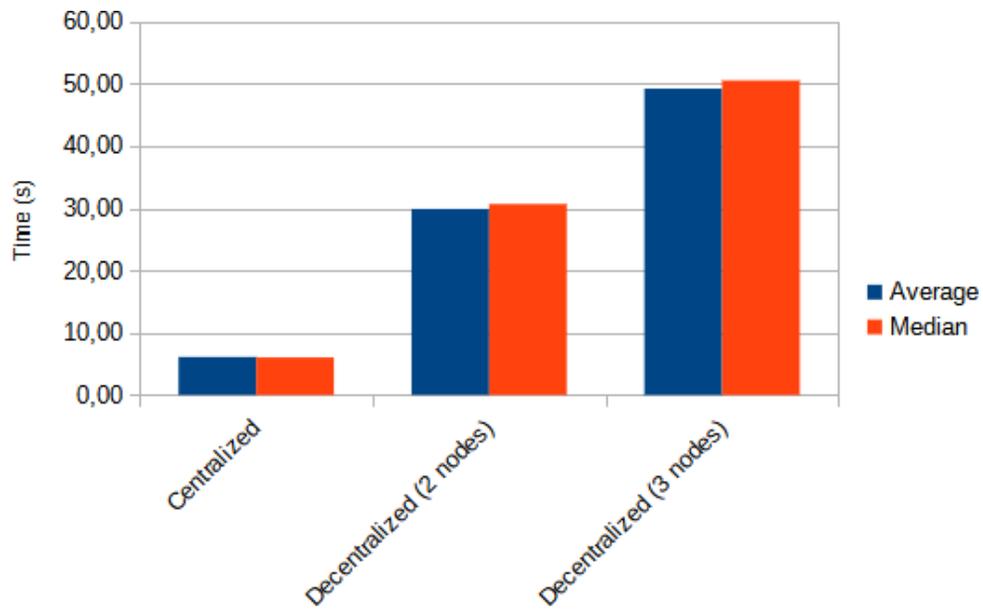

Figure 21: Response times for 24 KB files.

To carry out the quality tests, a total of 50 individual files were generated, each with a length of 1 million bits. Each of these files was subjected to the set of statistical tests specified in the NIST SP 800-22 standard.

Performing the tests on multiple independent bit files is aimed at minimizing the possibility that the results may be biased or influenced by the particularities of a single file or sequence. This practice allows researchers to detect possible inconsistencies or unwanted patterns under different conditions, providing a more complete and detailed view of the PRNG behavior. For the performed experiments, the Fortuna PRNG was initialized with bytes sourced from an operating system-specific entropy provider. Since the tests were conducted on a computer running Microsoft Windows, the entropy was derived from the BCryptGenRandom function, part of the Cryptography API: Next Generation (CNG) framework [82]. This function implements an algorithm for random number generation that adheres to the standards set forth in NIST SP 800-90.

In order to carry out a detailed comparison, the same tests were executed on a reference Fortuna implementation developed in Go by Jochen Voss [83]. In addition, the random number generation functions provided by the Python library "secrets" were subjected to the same tests. This library uses the random number generator provided by the underlying operating system, which is considered a reliable source of entropy. On Unix-based systems, "secrets" uses /dev/urandom to obtain entropy, while on Windows systems, it relies on the CryptGenRandom API to generate random numbers.

*6.4.2. Results*

The results of the comparison among the three evaluated PRNGs are summarized in Table 3. Such results demonstrate that HELENE implementation of Fortuna is highly effective for random number generation, with scores between 94% and 100%, outperforming J. Voss's Fortuna and Win-dows API in several tests. Moreover, it obtains perfect results (100%) in essential tests such as Frequency, Block Frequency and Serial, demonstrating a high quality of randomness. Its consistency, outstanding performance on key tests and robustness against more complex evaluations position it as a reliable high-quality solution. Although there are slight differences in some tests, these do not compromise its overall performance.

Table 3: NIST SP 800-22 result comparison.

| Test | HELENE Fortuna Accuracy Rate | J. Voss Fortuna Accuracy Rate | Windows API Accuracy Rate |
| --- | --- | --- | --- |
| Frequency | 100.0% | 100.0% | 96.0% |
| Block Frequency | 100.0% | 98.0% | 100.0% |
| Run | 98.0% | 98.0% | 98.0% |
| Longest Run of Ones | 98.0% | 100.0% | 100.0% |
| Binary Matrix Rank | 100.0% | 98.0% | 98.0% |
| Discrete Fourier Transform | 94.0% | 100.0% | 96.0% |
| Non-overlapping Template Matching | 98.0% | 100.0% | 100.0% |
| Overlapping Template Matching | 100.0% | 98.0% | 98.0% |
| Maurers Universal Statistical | 98.0% | 96.0% | 98.0% |
| Linear Complexity | 100.0% | 100.0% | 98.0% |
| Serial | 100.0% | 100.0% | 100.0% |
| Approximate Entropy | 98.0% | 98.0% | 98.0% |
| Cumulative Sums (Forward) | 98.0% | 100.0% | 96.0% |
| Cumulative Sums (Backward) | 100.0% | 100.0% | 98.0% |
| Random Excursion | 98.5% | 98.0% | 97.75% |
| Random Excursion Variant | 98.44% | 98.44% | 99.0% |

## 7. Conclusion

This article provided a detailed description of HELENE, a health service open-source platform supported by blockchain and that relies on a novel decentralized oracle. After analyzing the most relevant state of the art and background work, the communications architecture of the proposed system was presented together with the main subsystems and components, which include blockchain smart contracts based on Ethereum and the mentioned decentralized oracle. Then, the main processes managed by the platform were described, as well as the way lab tests were simulated to carry out experiments. Finally, a thorough description of the implementation of the platform components was provided, which was evaluated in terms of performance and security. The experimental results show that the average smart contract deployment time can be considered fast enough (roughly 13 s) and that batch transfers can accelerate significantly the operation of the network. Regarding the devised oracle, it can be concluded that the higher the number of decentralized Oracle Nodes, the higher the security, but the longer the time required by the oracle to respond. In addition, it was concluded that the developed PNRG used by the oracle complies with NIST SP 800-22 and is on par with the currently most popular implementations of Fortuna. Thus, this article, after an extensive description of the components and capabilities of HELENE, has shown the potential of the platform for delivering health services, thus setting the path for the creation of next generation DApps and for the future work of researchers, who can adapt HELENE to their needs, enhance it and evaluate its performance thanks to being open source.

**Funding**

This work has been funded by grant TED2021-129433A-C22 (HELENE) funded by MCIN/AEI/10.13039/501100011033 and the European Union NextGenerationEU/PRTR.

## Ethics statement

This study does not involve human participants, animal experiments, or any other procedures requiring ethical approval.